%% file: main.tex
\documentclass[11pt]{article}

\input{Macros}

\usepackage{graphicx}
\usepackage{amsmath}
\usepackage{caption}
\usepackage{subcaption}
\usepackage{epstopdf, epsfig}

\usepackage{bm}

\usepackage{mathabx}

\usepackage[font=scriptsize]{caption}

\begin{document}
\title{Time-Dependent Low-Rank Input-Output Operator for Forced Linearized Dynamics with Unsteady Base Flows}
\author[1]{Alireza Amiri-Margavi}
\author[1]{Hessam Babaee\thanks{Corresponding author. Email:h.babaee@pitt.edu.}}
\affil[1]{\footnotesize Department of Mechanical Engineering and Materials Science, University of Pittsburgh, 3700
O’Hara Street, Pittsburgh, PA 15213, USA}

\date{}
\maketitle

\begin{abstract}
Understanding the linear growth of disturbances due to external forcing is crucial for flow stability analysis, flow control, and uncertainty quantification. These applications typically require a large number of forward simulations of the forced linearized dynamics, often in a brute-force fashion.  When dealing with simple steady-state or periodic base flows, there exist powerful and cost-effective solution operator techniques. Once these solution operators are constructed, they can be used to determine the response to various forcings with negligible computational cost. However, these methods do not apply to problems with arbitrarily time-dependent base flows. This paper develops and investigates reduced-order modeling with time-dependent bases (TDBs) to build low-rank solution operators for forced linearized dynamics with arbitrarily time-dependent base flows. In particular, we use forced optimally time-dependent decomposition (f-OTD), which extracts the time-dependent correlated structures of the flow response to various excitations.
Several demonstrations are included to illustrate the utility of the f-OTD low-rank approximation for performing global transient stability analysis. Additionally, we demonstrate the application of f-OTD in computing the post-transient response of linearized Navier-Stokes equations to a large number of impulses, which has applications in flow control.
\\
\\
\textbf{Key words: } Unsteady base flows, low-rank approximation, time-dependent bases, forced optimally time-dependent decomposition  
\end{abstract}

\input{Introduction}
\input{Methodology}

\input{DemonstrationCases}

\input{Conclusions}

\input{Appendix}

\bibliographystyle{unsrt}
\bibliography{HB,AR}
\end{document}

%% file: Macros.tex
\usepackage{amsmath}
\usepackage{amsfonts}
\usepackage{pgfplots}
\usepackage{amssymb}
\usepackage{tikz}
\usetikzlibrary{positioning}
\usetikzlibrary{matrix, arrows,decorations.pathmorphing}
\usetikzlibrary{patterns}
\usetikzlibrary{through,calc,3d}
\usetikzlibrary{plotmarks}
\usepackage[titletoc]{appendix}
\usepackage[T1]{fontenc}
\usepackage[numbers,square,sort&compress]{natbib}
\usepackage{subfigure}                  
\usepackage[affil-it]{authblk}
\usepackage{algorithm}
\usepackage{algpseudocode}

\newcommand{\inner}[2]{\big< #1 ,  #2 \big>}
\newcommand{\bm}[1]{\ensuremath{\mathbf{#1}}}

%% file: Introduction.tex
\section{Introduction}

Global stability analysis aims to provide a quantitative description of flow behavior by analyzing infinitesimal external disturbances superimposed on a complex base flow (\cite{schmid2007nonmodal}). The external disturbances may represent acoustic waves and vibration (\cite{qin2016response}), surface roughness (\cite{schrader2009receptivity}), free-stream turbulence (\cite{schneider2001effects, joo2012continuous,brinkerhoff2015numerical}), or neglected nonlinear terms in the governing equations (\cite{delsole1995stochastically}). The disturbances may also be imposed on the initial condition.

Computational cost is one of the underlying challenges of performing global linear stability analysis (\cite{KLC23}). Many questions in global linear stability analysis require computing the response of the flow to high-dimensional infinitesimal disturbances. This means the number of independent disturbances that need to be considered is typically very large and often proportional to the size of the state vector representing the flow in the discrete form. For example, building an input-output operator in a brute-force manner requires solving the inhomogeneous linearized system for the external excitation imposed on each state variable. Another example is determining the most amplified forcings and the most receptive states, which necessitates solving an optimization problem involving many expensive numerical simulations of the linearized Navier-Stokes equations. However, the computational cost greatly varies depending on the complexity of the flow, whether the flow is homogeneous in any physical dimension, and whether the flow is steady, periodic, or arbitrarily time-dependent.

There are efficient numerical algorithms for performing global stability analysis when the base flow is steady. For steady-state base flows, modal global stability analysis requires solving large-scale eigenvalue problems. Krylov-based techniques (\cite{edwards1994krylov, frantz2023krylov}) have demonstrated excellent performance in extracting leading eigenmodes by projecting the linearized dynamics onto a lower-dimensional subspace, obtained through the orthonormalization of snapshots of the flow field. This technique has been used for the global stability analysis of a three-dimensional jet in crossflow (\cite{bagheri2009global}), where the base flow is forced to a steady-state solution using the selected frequency damping technique (\cite{aakervik2006steady}). On the other hand, performing nonmodal global stability analysis, even when the base flow is steady, requires computing a very large number of eigenmodes, especially for highly nonnormal linearized operators.  Although the fundamental solution operator formalism can be employed for stability analysis, computing this operator is computationally prohibitive for most practical flows and is fraught with numerical challenges (\cite[Section 6.4.2]{Henningson_Schmidt}).

Similarly,  computational algorithms based on resolvent analysis (\cite{trefethen1993hydrodynamic}) have been successfully employed for computing the response to harmonic forcings when the base flow is steady (\cite{schmidt2018spectral, bae2020resolvent, skene2022sparsifying, cook2022free}).  In resolvent analysis, the linear operator is obtained by linearizing the Navier-Stokes equations around the steady-state base flow or time-averaged mean flow (\cite{mckeon2010critical}). For many flows, the resolvent operator has fast decaying singular values, which allows accurate low-rank approximation. This enables the development of numerous diagnostic and predictive tools including the extraction of dominant coherent structures in turbulent flows (\cite{abreu2020resolvent}),  guiding the placement of sensors and actuators for flow control (\cite{martini2022resolvent, jin2022optimal}), and reduced-order modeling of turbulent flows (\cite{mckeon2017engine, sun2020resolvent}).

The computational cost of constructing the resolvent operator and computing its low-rank approximation can be prohibitive for complex flows. If the flow has more than one non-homogeneous direction, the explicit construction of the resolvent operator becomes costly or infeasible due to the storage and inversion of large matrices. Several methodologies have been proposed to address these issues, including the use of a modified randomized singular value decomposition (RSVD) (\cite{ribeiro2020randomized}) and methods based on time-stepping (\cite{MAB10, MRTC21}) that involve the time integration of the linearized Navier-Stokes equation and its adjoint.

When the base flow is harmonically time-dependent, there are several approaches to perform stability analysis in a cost-effective manner. When the base flow is assumed to be time-invariant, the interactions across various frequencies cannot be computed. Floquet theory enables the stability analysis of linear systems that have periodic time dependence (\cite{L2003, KLH23}). Resolvent analysis has also been recently extended to base flows with periodic time dependence (\cite{padovan2020analysis}), referred to as harmonic resolvent analysis. The computational cost of performing harmonic resolvent analysis can be significant as the number of considered frequencies increases. To this end, a methodology based on time-stepping has recently been proposed to address this issue (\cite{FJBT24}). To identify the optimal forcing and response modes that exhibit sparsity both spatially and temporally, a novel variant of resolvent analysis has been introduced (\cite{lopez2023sparsity}). This method, termed space-time resolvent analysis, has been validated through application to statistically stationary channel flow and a time-periodic Stokes boundary layer.

The ability to simulate the evolution of disturbances for arbitrarily time-dependent flows in a cost-effective manner has significant implications for the stability analysis of complex flow problems. When the base flow is unsteady, the linearized operator continuously varies with time, and in general, the effect of this variation must be taken into account in the stability analysis. As mentioned above, while there are well-established and computationally efficient tools for performing linear stability analysis for problems with steady-state and time-periodic base flows, there are far fewer computationally efficient techniques for the stability analysis of unsteady flows (\cite[Section 6.4.2]{Henningson_Schmidt}). Although the fundamental solution operator formalism can be employed for stability analysis, computing this operator is computationally prohibitive for most practical flows. 

An example application was recently studied by \cite{KNH24}, where linear stability analysis was performed for an aerofoil undergoing small-amplitude pitching motion. In this analysis, the base flow undergoes deformation due to a global instability mechanism, leading to secondary instability followed by a rapid breakdown to turbulence. \cite{KNH24} investigates the effect of perturbation in the initial condition, which can be expressed as an initial-value problem for a homogeneous time-variant linear system. However, the effect of disturbance at the upstream boundary conditions, surface roughness, or neglected nonlinear terms can be formulated as a time-variant linear system subject to high-dimensional forcing. In these types of problems, the base flow variation with time is non-periodic and arbitrary, and it is not clear what instant of base flow should be chosen for using classical stability analysis methods such as eigenvalue decomposition. This is particularly important for flows with transient instabilities, where a steady base flow cannot be assumed, and using a time-averaged base flow results in a significant loss of information.

More broadly, the analysis of transient instabilities in high-dimensional dynamical systems such as turbulent flows demands cost-effective techniques to track the evolution of high-dimensional disturbance space. Transient instabilities are finite-time events, such as turbulent bursts or extreme events, whose response is markedly different from the high-dimensional attractor where the dynamical system spends most of its lifetime (\cite{T21}). Examples include turbulent bursts in laminar or turbulent flows (\cite{AML11, CD13, FCRP17, SS19}), the transition between a chaotic attractor and non-chaotic traveling waves (\cite{MCT11}), and atmospheric blocking events (\cite{DCG12}).

Our objective is to develop a low-rank approximation of the solution operator for arbitrarily time-dependent base flows.  This involves working with the arbitrarily time-dependent matrix obtained by linearizing the Navier-Stokes equations around the instantaneous base flow.

Our approach is to develop an input-output low-rank approximation based on forced optimally time-dependent (f-OTD)  (\cite{donello2022computing}) decomposition. The f-OTD decomposition is an extension of OTD approximation  (\cite{babaee2016minimization}) and it was originally developed by  \cite{donello2022computing} for computing sensitivities in dynamical systems. The OTD and f-OTD decompositions have been utilized for many applications including the computation of finite-time Lyapunov exponents (\cite{babaee2017reduced}), detection of the edge of chaos in dynamical systems (\cite{beneitez2020edge}),  prediction of bursting phenomena in high-dimensional dynamical systems (\cite{FS16}), global transient stability analysis of a time-dependent base flow over an airfoil  (\cite{KNH24}), control and linear instabilities (\cite{blanchard2019stabilization,kern2021transient}),  and skeletal kinetics (\cite{nouri2022skeletal, nouri2024skeletal}).

The OTD and f-OTD belong to a wider class of low-rank approximation methods used for matrix differential equations (MDEs) that rely on time-dependent bases (TDBs), where the solution of the MDE is evolved on a manifold of low-rank matrices. These TDB-based approximations originated in quantum chemistry and were utilized to solve high-dimensional Schrödinger equation (\cite{Beck:2000aa}). In (\cite{KL07}), such techniques were presented for general MDEs under the name of dynamical low-rank approximation (DLRA). The dynamically orthogonal (DO) decomposition is another TDB-based method, tailored for solving stochastic partial differential equations (\cite{SL09}). Both DO and f-OTD techniques adopt a two-matrix factorization ($\mathbf{V} \approx \mathbf{U} \mathbf{Y}^\mathrm{T}$) for the full-rank matrix, while DLRA leverages a three-matrix factorization ($\mathbf{V} \approx \mathbf{U} \bm{\Sigma} \mathbf{Z}^\mathrm{T}$). See Section \ref{Methodology} for the notation used here. As demonstrated in \cite[Theorem 2.2]{PB20}, two-matrix and three-matrix factorization methods are equivalent. TDB-based approximations have been applied to various applications such as turbulent combustion (\cite{RNB21}) and kinetics equations (\cite{einkemmer2018low, KS23}).

There are several other low-rank approximation techniques based on time-evolving subspaces. Spectral Proper Orthogonal Decomposition (SPOD) extracts orthonormal time-dependent POD modes, with each mode oscillating at a single frequency (\cite{TSC18}). Recently,  \cite{FT23}  proposed space-time POD where the modes vary arbitrarily with time. Both SPOD and space-time POD contrast with space-only POD, which is based on static subspaces. SPOD and space-time POD follow the same mathematical formalism as space-only POD, with the main difference being that the inner product of the SPOD and space-time modes is taken with respect to an integral over space and time.

The SPOD, space-time POD, and f-OTD low-rank approximations extract different types of correlations: SPOD and space-time POD extract spatiotemporal coherent structures, whereas f-OTD extracts instantaneous same-time correlations between different samples of the linearized dynamics. These different samples might be generated due to varying initial conditions or different forcings. Consequently, f-OTD modes are instantaneously orthonormal and vary arbitrarily with time, similar to space-time POD, rather than being time-periodic like SPOD. Another difference is that, in SPOD and space-time POD, the bases are time-dependent while the coefficients of the bases are constant; in f-OTD, both the bases and the coefficients are time-dependent. Additionally, f-OTD does not assume statistical stationarity, allowing it to be applied to flows that are not statistically stationary.

There is also an important difference between f-OTD and other data-driven low-rank approximation approaches: The f-OTD is an equation-based low-rank approximation. Unlike SPOD and space-time POD, where modes are computed from data, evolution equations for the f-OTD subspace are derived from the linearized Navier-Stokes equations and no offline data generation is required. 

Perhaps it is more appropriate to compare SPOD and space-time POD to the DO decomposition (\cite{SL09}), which, in addition to being nonlinear and stochastic, shares all the attributes of f-OTD.

Recently, a continuous-time balanced truncation (CTBT) (\cite{PR24}) has been proposed for time-periodic flows. This methodology extends the balanced POD—introduced by \cite{R05} for balanced truncation of time-invariant systems—to time-periodic flows. The approach presented by \cite{PR24} results in a low-rank approximation based on time-evolving subspaces, where a time-continuous coordinate change and truncation are achieved by setting the time-dependent reachability and observability Gramians to be equal and diagonal. The CTBT low-rank approximation is closely related to harmonic resolvent analysis. 

The similarity between f-OTD and CTBT is that the end product for both methodologies is a low-rank approximation based on time-evolving subspaces for forced time-dependent linear systems. However, the goals of these low-rank approximations are different. CTBT aims to obtain a balanced truncation, which requires computing the Gramians. In contrast, f-OTD seeks to approximate the response of the forced linearized dynamics and does not aim for a balanced truncation; thus, computing Gramians is not needed in f-OTD. In that sense, f-OTD is more closely related to harmonic resolvent analysis than to CTBT. Another important distinction is that the f-OTD evolution equations are solved in the time domain rather than the frequency domain. Therefore, f-OTD can be utilized for forced linear systems that are arbitrarily time-dependent. In this paper, we consider one such problem where we use f-OTD to compute the response of linearized Navier-Stokes equations to a large number of impulses.

The structure of this paper is as follows: In Section \ref{Methodology}, we develop the f-OTD methodology for constructing low-rank solution operators for the forced linearized dynamics. In Section \ref{Demonstration}, we demonstrate the capability of f-OTD on several examples, including a toy model, 1D Burgers equation, a two-dimensional temporally evolving jet, the chaotic Kolmogorov flow, and two-dimensional decaying isotropic turbulence. Finally, we discuss the main findings and provide concluding remarks in Section \ref{Conclusion}.

%% file: Methodology.tex
\section{Methodology\label{Methodology}}

\subsection{Matrix differential equations for the operator evolution} 
In this section, we formulate the evolution of the disturbance due to different external forcings as an MDE. To this end, we consider the semi-discretized system where the governing equations are discretized in space. We consider a nonlinear  dynamical system in the form of:
\begin{equation}\label{eq:state}
\dot{\mathbf{u}}= \mathbf{g}(\mathbf{u},t),   \quad t \in I=[0,T],
\end{equation}
where  $\mathbf{u}(t) \in \mathbb{R}^n$  represents the state of the dynamical system and  $\mathbf{g}(\mathbf{u},t) \in  \mathbb{R}^n$ is a nonlinear function of the state. We denote  the solution of Eq. \ref{eq:state} with the initial condition
$\mathbf{u}(t=0)=\mathbf{u}_0$ and we denote the state of the trajectory at time $t$  by $\mathbf{u}(t)$. 
We consider the problem of the nonlinear dynamics prescribed by Eq.~\ref{eq:state} subject to infinitesimal forcing. The resulting disturbance   $\mathbf{v}$ satisfies the linear equation of variations:
\begin{equation}\label{eq:state_2}
 \dot{\mathbf{v}}= \mathbf{L}\big(\mathbf{u},t\big) \mathbf{v} + \mathbf{f}, \quad \mathbf{v}(0)=\mathbf{0},
\end{equation}
where, $\mathbf{v}(t) \in \mathbb{R}^n$ is the state of the disturbance,  $\mathbf{L}(\mathbf{u},t) \in \mathbb{R}^{n\times n}$, is the matrix of the Jacobin of the vector field   $\mathbf{g}(\mathbf{u},t)$, i.e.,  $\mathbf{L}_{ij}=\partial \mathbf{g}_i/\partial \mathbf{u}_j, i,j=1, \dots, n $, and  $\mathbf{f}(t) \in \mathbb{R}^n$ is the forcing vector. Eq.~\ref{eq:state_2} describes the effect of external forcing. The external forcing can represent free-stream turbulence, wall roughness, or neglected terms, such as nonlinear terms (\cite{schmid2007nonmodal}).  We consider cases where $\mathbf{f}$ belongs to a class of forcing: $\mathbf{f} \in \mathcal{S}$ with the dimension $d$, i.e., $d=\mbox{dim}(\mathcal{S})$. Therefore, any forcing in the subset $\mathcal{S}$ can be determined with a set of $d$ coordinates and a set of  basis vectors  for $\mathcal{S}$:
\begin{equation}\label{eq:force_coord}
    \mathbf{f} = \sum_{j=1}^d y_j \mathbf{f}_j, \quad \forall \mathbf{f} \in \mathcal{S},
\end{equation}
where $\mathbf{f}_j  \in \mathbb{R}^n$ constitute a set of  basis vectors for $\mathcal{S}$, i.e.,  $\mathcal{S}=\mbox{span}(\mathbf{f}_1, \mathbf{f}_2, \dots ,\mathbf{f}_d)$ and   $ y_j \in \mathbb{R}$ are the coordinates.  In the presented formulation, the forcing basis vectors do not need to be orthogonal; they only need to be independent. To investigate the response of Eq.~\ref{eq:state_2} for any $\mathbf{f} \in \mathcal{S}$ we cast Eq.~\ref{eq:state_2} as an MDE:
\begin{equation}\label{eq:MDE_FOM}
 \dot{\mathbf{V}}= \mathbf{L}\big(\mathbf{u},t\big) \mathbf{V} + \mathbf{F}, \quad \mathbf{V}(0) = \mathbf{0},
\end{equation}
where $\mathbf{F} =[\mathbf{f}_1, \mathbf{f}_2, \dots ,\mathbf{f}_d] \in \mathbb{R}^{n\times d}$ is the matrix of   forcing basis vectors and  $\mathbf{V} =[\mathbf{v}_1, \mathbf{v}_2, \dots ,\mathbf{v}_d] \in \mathbb{R}^{n\times d}$ is the matrix of corresponding response. The forcing matrix $\mathbf F$
can be the actual forcing matrix as long as the forcings are independent. The dependent forcings need not be considered, as the response of the linear system to those dependent forcings can be obtained as a linear combination of the responses to the independent forcings.

We refer to Eq.~\ref{eq:MDE_FOM} as the full-order model (FOM). Any column of $\mathbf{V}(t)$  contains the disturbances of all state variables for a single forcing and each row of $\mathbf{V}(t)$  contains the disturbances of a single state variable for all forcings. Each column of MDE~\ref{eq:MDE_FOM} can be solved independently of other columns while each row of MDE~\ref{eq:MDE_FOM} is dependent to other rows and therefore each row cannot be solved independently. 

The solution of FOM given by Eq.~\ref{eq:MDE_FOM} can be written as:
\begin{equation}
    \mathbf{V}(t) = \int_0^t \Phi_{\tau}^t \mathbf F(\tau) d\tau,
\end{equation}
where $\Phi_{\tau}^t \in \mathbb{R}^{n \times n}$ is the fundamental solution operator  and its evolution is governed by the homogeneous disturbance evolution equation given by:
\begin{equation}
\dot{\Phi}_{\tau}^t = \mathbf{L}\big(\mathbf{u},t\big) \Phi_{\tau}^t, \quad \Phi_{\tau}^{\tau}= \mathbf{I}, \quad t \geq 0,
\end{equation}
where $\mathbf{I} \in \mathbb{R}^{n \times n}$ is the identity matrix. Since the MDE given by Eq.~\ref{eq:MDE_FOM} is linear, the disturbance state for any $\mathbf{f} \in \mathcal{S}$ is given by:
\begin{equation}
\mathbf{v}(t) = \big (\int_0^t \Phi_{\tau}^t \mathbf F(\tau) d\tau \big ) \mathbf{y}, 
\end{equation}
where $\mathbf{y} = [y_1,y_2,\dots,y_d]^\mathrm{T} \in \mathbb{R}^d$ is the coordinate of forcing $\mathbf{f}$ in the basis $\mathbf{F} = [ \mathbf{f}_1, \mathbf{f}_2, \dots, \mathbf{f}_d]$. 
We define the operator $\mathbf{H}^t_\mathcal{S}$ as  follows:
\begin{equation}\label{eq:operator_full-rank}
    \mathbf{H}^t_\mathcal{S}( \cdot) =  \big (\int_0^t \Phi_{\tau}^t \mathbf F(\tau) d\tau \big ) \big(\cdot \big), 
\end{equation}
where $\mathbf{H}^t_{\mathcal{S}}(\cdot): \mathbb{R}^d  \rightarrow \mathbb{R}^{n}$. 
To calculate the response of any forcing $\mathbf{f} \in \mathcal{S}$ using the solution operator $\mathbf{H}^t_{\mathcal{S}}$, $\mathbf{f}$ must be first expressed in forcing basis coordinate system according to Eq.~\ref{eq:force_coord}. This is expressed as: $\mathbf{f} = \mathbf{F} \mathbf{y}$. 
 The coordinate vector $\mathbf{y}$ is the input to the operator $\mathbf{H}^t_{\mathcal{S}}(\cdot)$.   The output of the operator $\mathbf{H}^t_{\mathcal{S}}$ is the state of the disturbance, i.e., $\mathbf{v}(t) = \mathbf{H}^t_{\mathcal{S}}(\mathbf{y})$. The operator $\mathbf{H}^t_{\mathcal{S}}$ is an instantaneously linear operator that varies smoothly with time to adapt in response to changes in the state of the disturbances. 
 In this view, the operator $\mathbf{H}^t_\mathcal{S}$ is a time-dependent matrix, and computing this operator is equivalent to computing   $\mathbf{V}(t)$. 
However, solving the FOM  is computationally prohibitive, since both  $n$ and $d$ are very large for most practical fluid dynamics applications.

\subsection{On-the fly low-rank approximation  with f-OTD\label{MethodB}}
We present a novel application of low-rank approximation for MDE based on time-dependent subspaces to approximate the solution of the FOM given by MDE \ref{eq:MDE_FOM}. In particular, we formulate a low-rank approximation based on f-OTD. The f-OTD formulation seeks to approximate the solution of the FOM with a rank-$r$ matrix:
\begin{equation}\label{eq:state_7}
\mathbf{V}(t) \approx \mathbf{U}(t) \mathbf{Y}(t)^\mathrm{T},
\end{equation}  
where $\mathbf{U}(t) =[\mathbf{u}_1(t), \mathbf{u}_2(t),  \dots,  \mathbf{u}_r(t)] \in \mathbb{R}^{n \times r}$ is  matrix of the f-OTD modes
and $\mathbf{Y}(t)=[\mathbf{y}_1(t), \mathbf{y}_2(t), \dots,   \mathbf{y}_r(t)]\in \mathbb{R}^{d \times r}$ is the matrix of  f-OTD coefficient. The f-OTD modes are instantaneously orthonormal, i.e, ${\mathbf{u}_i(t)}^\mathrm{T} {\mathbf{u}_j(t)}= \delta_{ij}$ and they represent a low-rank time-dependent basis for the columns of matrix $\mathbf{V}(t)$.  The matrix  $\mathbf{Y}(t)$ represents a low-rank time-dependent basis for the rows of matrix $\mathbf{V}(t)$. In the f-OTD formulation, the columns of $\mathbf{Y}(t)$ are not orthogonal to each other. 

Because the f-OTD is a low-rank approximation, it cannot satisfy the  FOM (Eq.~\ref{eq:MDE_FOM}). Therefore, the f-OTD expansion satisfies the MDE~\ref{eq:MDE_FOM} with a residual. To this end, a variational  principle is utilized, which  seeks to minimize this residual by optimally evolving the f-OTD components ${ \mathbf{U}(t)}$ and  ${\mathbf{Y}(t)}$ as shown below
\begin{equation}\label{eq:state_8}
\mathcal{F}\big( \dot{\mathbf{U}}, \dot{\mathbf{Y}}\big)= 
\bigg \| \frac{d\big(\mathbf{U}{\mathbf{Y}}^\mathrm{T}\big)}{d t} -\mathbf{L}\mathbf{U} \mathbf{Y}^\mathrm{T} -\mathbf{F} \bigg \|^{2}_{F},
\end{equation}
where $\| \cdot\|_F$ denotes the  Frobenius norm. The dependence of matrices on $t$ and $\mathbf{u}$ is dropped for brevity. 
The above variational principle is subject to the orthonormality constraints of the spatial modes, i.e., ${\mathbf{U}}^\mathrm{T}{\mathbf{U}}=\mathbf{I}$. The control parameters for the above function are $\dot{\mathbf{U}}$ and  $\dot{\mathbf{Y}}$.   The orthonormality constraints can be enforced via the Lagrange multipliers. To this end, we first take a time derivative of the orthonormality constraints:
\begin{equation}
\dot{\mathbf{u}}_i^\mathrm{T}{\mathbf{u}_j}+{\mathbf{u}_i}^\mathrm{T}{\dot{\mathbf{u}}_j}=0.
\end{equation}
We denote  $\Psi_{ij}: = {\mathbf{u}_i}^T\dot{\mathbf{u}}_j$. It is clear from the above equation that  $\boldsymbol{\Psi}=[\mathbf{\psi}_{ij}]\in  \mathbb{R}^{ r \times r}$ is an arbitrary skew-symmetric matrix, $\big(\boldsymbol{\Psi}^\mathrm{T}=-\boldsymbol{\Psi}\big)$. Incorporating the orthonormality condition into the variational principle leads to an unconstrained optimization problem as follows:
\begin{equation}\label{Eqn:min_princ}
\begin{aligned}
 \mathbf{\mathcal{G}}(\dot{\mathbf{U}}, \dot{\mathbf{Y}},\boldsymbol \lambda)  &= \left\Vert \frac{d (\mathbf{U} \mathbf{Y}^\mathrm{T})}{d t}  -\mathbf{L}\mathbf{U} \mathbf{Y}^\mathrm{T} -\mathbf{F}   \right\Vert_F^{2}+ \sum_{i,j=1}^r \lambda_{ij}  \big( \mathbf{u}_i^\mathrm{T}\dot{\mathbf{u}}_j - \Psi_{ij} \big),
\end{aligned}
\end{equation}
where $\boldsymbol \lambda=[\lambda_{ij}]$ with $ i,j=1, \dots, r$ are  Lagrange multipliers.  The procedure for derivation of the optimality conditions is provided in (\cite{donello2022computing}). From the first-order optimality conditions of the unconstrained functional, the closed-form evolution equation for $\mathbf{U}$ and  $\mathbf{Y}$ are derived. These equations are:
\begin{align}
\dot{\mathbf{U}} &= \mathbf{L}\mathbf{U}-\mathbf{U}\mathbf{L}_r+\big(\mathbf{F}\mathbf{Y}-\mathbf{U}\mathbf{U}^\mathrm{T}\mathbf{F}\mathbf{Y}\big){\mathbf{C}}^{-1}, \quad
   \mathbf{U}(t_0)=\mathbf{U}_0, \label{eq:u_ev_DO}\\
\dot{\mathbf{Y}}&= \mathbf{Y}\mathbf{L}_r^\mathrm{T}+\mathbf{F}^\mathrm{T}\mathbf{U}, \quad
   \mathbf{Y}(t_0)=\mathbf{Y}_0, 
\label{eq:y_ev_DO}
\end{align}
where $\mathbf{C} = \mathbf{Y}^\mathrm{T} \mathbf{Y}\in \mathbb{R}^{  r \times r}$ is the reduced correlation matrix, and $\mathbf{L}_r=\mathbf{U}^\mathrm{T}{\mathbf{L}\mathbf{U}} \in \mathbb{R}^{  r \times r} $ is the reduced linear operator. We refer to f-OTD as an ``on-the-fly" low-rank approximation because it does not require the offline data generation commonly needed in data-driven reduced-order modeling. Instead, the matrix $\mathbf{V}(t)$ is solved directly in its low-rank form, i.e., by evolving $\mathbf{U}(t)$ and $\mathbf{Y}(t)$.  The f-OTD subspaces are initialized by first solving the FOM for one $t_0 = \Delta t$ and then computing the first $r$ singular vectors of  $\mathbf{V}(t_0) \approx \mathbf{U}(t_0) \bm{\Sigma}(t_0) \mathbf{Z}^\mathrm{T}(t_0)$, where $\mathbf{U}(t_0) \in \mathbb{R}^{n\times r}$ is used as the initial condition for the f-OTD modes and  $\mathbf{Y}(t_0) = \mathbf{Z}(t_0) \bm{\Sigma}(t_0)$, where $\bm{\Sigma}(t_0) \in \mathbb{R}^{r\times r}$ and  $\mathbf{Z}(t_0) \in \mathbb{R}^{d\times r}$ are the matrices of singular values and right singular vectors, respectively. In the case of very large-scale dynamical systems, one can determine the initial conditions for the f-OTD matrices by a targeted sparse sampling of the FOM, which decreases the computational costs (\cite{DPNFB23}). This approach is based on the CUR low-rank approximation, where the approximation error can be made arbitrarily close to that of the same-rank SVD low-rank approximation through oversampling.

One way to interpret the f-OTD evolution equations is the following: Eq.~\ref{eq:u_ev_DO} determines the evolution of a time-dependent subspace in the phase space of the dynamics and Eq.~\ref{eq:y_ev_DO} is the ROM obtained by the orthogonal projection of the FOM onto the time-evolving subspace $\mathbf{U}$. The above constrained minimization problem can be alternatively solved using Riemannian optimization, where the solution of MDE~\ref{eq:MDE_FOM} is constrained on a rank-$r$ manifold. This will lead to the dynamical low-rank approximation for MDEs (\cite{KL07}), which results in an equivalent low-rank approximation to the f-OTD formulation.   

The evolution Eqs.~\ref{eq:u_ev_DO} \& \ref{eq:y_ev_DO} are presented for generic dynamical systems. The f-OTD evolution equations for the incompressible Naiver-Stokes equations are derived in Appendix \ref{Appendix_f-OTD_NS}. 

Note that $\mathbf{L}_r$ is time-dependent, even when $\mathbf{L}$ is not, due to the time-dependence of $\mathbf{U}$. The OTD equations can be derived from the f-OTD equations by setting $\mathbf{F}$ to $\mathbf{0}$. As demonstrated in (\cite{babaee2016minimization}), for a homogeneous linearized flow (where $\mathbf{F}=\mathbf{0}$) with a constant $\mathbf{L}$, the subspace $\mathbf{U}$ initially captures the nonnormal growth in early disturbance evolution and eventually converges asymptotically to the rank-$r$ subspace spanned by the $r$ most unstable eigenvectors of $\mathbf{L}$. Furthermore, as shown in (\cite{babaee2017reduced}), when $\mathbf{L}$ is time-variant, the subspace $\mathbf{U}$ converges exponentially to the eigenvectors of the Cauchy–Green strain tensor that corresponds to the most intense finite-time instabilities.

{The rank of the f-OTD approximation is a hyperparameter and increasing the rank improves the accuracy. In fact, for $r=d$, the f-OTD equations recover the FOM. One heuristic approach to determine the rank is to  find the smallest $r$ such that the ratio of the last resolved singular value to the Frobenius norm of the operator given by
\begin{equation*}
    \epsilon(t) = \frac{\sigma_r(t)}{\big(\sum_{i=1}^r \sigma_i^2(t)\big )^{1/2}}
\end{equation*}
is smaller than a threshold value.  
Ideally, the f-OTD should be rank adaptive because, for a given accuracy criterion, the rank of the approximation needs to vary with time as the singular values of the solution matrix are time-dependent. This criterion has been recently used for a rank-adaptive TDB-based low-rank approximation for MDEs (\cite{DPNFB23}) and tensor differential equations (\cite{GB24-TT}). There are also other approaches for rank adaptivity for TDB-based low-rank approximations; see, for example, (\cite{DRV21, CKL22}). All of the simulations performed in this paper are based on a fixed rank.

The f-OTD evolution equations become stiff when $\mathbf C$ is ill-conditioned as the inversion of this matrix appears in Eq. \ref{eq:u_ev_DO}. A similar issue also occurs for DLRA and other equivalent TDB-based low-rank approximations. This issue has been the subject of research and several different algorithms have been proposed that are robust in the presence of near singular or rank-deficient correlation matrix or singular value matrix; see for example  \cite{LO14,CL21,DPNFB23}.

\subsubsection{Computational cost}\label{sec:comp_cost}
In the practical applications,   $n \sim \mathcal{O}(10^9)$ and $d \sim \mathcal{O}(10^3)$-$\mathcal{O}(10^9)$  and therefore,  solving the FOM is computationally prohibitive due to input/output (I/O)  and memory requirements as well as the floating point operations costs, which scales at least with $O(nd)$. Eq.~\ref{eq:u_ev_DO} can be simplified to: $\dot{\mathbf{u}}_i = \mathbf{L}\mathbf{u}_i + \mathbf{r}_i, \ i=1,\dots, r$.  The computational cost of evolving this equation is the same as solving $r$ samples of Eq. \ref{eq:state_2}, i.e., $\mathcal{O}(rn)$.    Eq.~\ref{eq:u_ev_DO} is a thin MDE and the computation cost of solving this equation is $\mathcal{O}(rs)$.  The right-hand side of the f-OTD evolution equations requires computing $\mathbf{U}^\mathrm{T}\mathbf{F}$ and $\mathbf{F}\mathbf{Y}$, which are both $\mathcal{O}(rdn)$. If the forcing is sparse, for example, when the forcing is localized at the boundaries, this cost can become negligible. When $\mathbf{F}$ is a dense matrix, the cost of computing these matrix-matrix multiplications can be reduced to $\mathcal{O}(r(n+d))$ using oblique projections and sparse sampling (\cite{NB23, DPNFB23}). 

The OTD low-rank approximation ($\mathbf F = \mathbf 0$) has been used for stability analysis of three-dimensional flow including a three-dimensional jet in a cross flow (\cite{babaee2016minimization}) and three-dimensional time-dependent flow around an aerofoil (\cite{KNH24}).

\subsubsection{Mode ranking and the canonical form}\label{sec:mode_rank}

The f-OTD subspace captures the most dominant perturbation subspace, however, the f-OTD modes and their coefficients are not ranked energetically. For example, $\{\mathbf u_1,\mathbf y_1\}$ are not aligned with the first left and right singular vectors of the operator. 

The f-OTD modes and the forcing coefficient can be ranked based on their significance after they are computed via Eqs. \ref{eq:u_ev_DO} and \ref{eq:y_ev_DO}.  This is achieved through the eigen-decomposition of the reduced correlation matrix:
$\mathbf{C}(t)\mathbf{R}(t)=\mathbf{R}(t)\bm{\Lambda}(t)$,
where $\mathbf C(t) = \mathbf Y(t)^\mathrm{T} \mathbf Y(t)$ and  the matrix $\mathbf{R}(t) \in \mathbb{R}^{r\times r}$ and $\bm{\Lambda}(t) = \mbox{diag}(\lambda_1(t), \dots, \lambda_r(t))$ are  the eigenvector and eigenvalues  of $ \mathbf{C}(t)$ matrix, respectively.   The eigenvalues of the correlation matrix are nonnegative, i.e.,  $\lambda_i(t)\geq 0$  since $ \mathbf{C}$ is a positive symmetric matrix and $ \mathbf{R}(t)$  is an orthonormal matrix. The eigenvalues of  $\mathbf{C}(t)$ can be ranked such that ${\lambda}_1(t) \geq {\lambda}_2(t) \geq ... \geq  {\lambda}_r(t) \geq 0$. It is straightforward to show that $\sigma_i(t) = \lambda_i^{1/2}(t)$ are the singular values of the matrix $\mathbf{U}(t) \mathbf{Y}(t)^\mathrm{T}$. The f-OTD modes and modal coefficients can be rotated by disturbance  energy (${\sigma}_{i}^2$) in descending order as follows:
\begin{equation}
    \mathbf{\tilde {Y}}(t) = \mathbf{Y}(t) \mathbf{R}(t) \bm{\Sigma}^{-1}(t), \quad \quad \mathbf{\tilde {U}}(t)= \mathbf{U}(t)\mathbf{R}(t).
\end{equation}
where $\bm{\Sigma}(t) =\mbox{diag}(\sigma_1(t), \dots, \sigma_r(t))$ is the matrix of singular values,  $\mathbf{\tilde {Y}}(t)$ and $\mathbf{\tilde {U}}(t)$ are referred to as the bio-orthonormal (BO) form of the reduction, i.e., $\mathbf{\tilde {U}}(t)^\mathrm{T}\mathbf{\tilde {U}}(t) = \mathbf{I}$ and $\mathbf{\tilde {Y}}(t)^\mathrm{T}\mathbf{\tilde {Y}}(t) = \mathbf{I}$, where $\mathbf{I} \in \mathbb{R}^{r\times r}$ is the identity matrix. We note that the BO form \{$\mathbf{\tilde {Y}}(t)$, $\bm{\Sigma}(t)$,$\mathbf{\tilde {U}}(t)$\} and \{$\mathbf{U}(t)$ ,$\mathbf{Y}(t)$\}  are equivalent, i.e., 
$\mathbf{\tilde {U}}(t)\bm{\Sigma}(t){\mathbf{\tilde {Y}}}(t)^\mathrm{T}={\mathbf{U}(t)}{\mathbf{Y}}(t)^\mathrm{T}$.

It is possible to write the evolution equations of the f-OTD modes directly in the energetically ranked form, i.e., the evolution equations for ${\tilde{\mathbf{U}},\tilde{\mathbf{Y}}\boldsymbol \Sigma }$ or ${\tilde{\mathbf{U}}\boldsymbol \Sigma,\tilde{\mathbf{Y}} }$. These evolution equations would result in equivalent subspaces to those obtained by solving Eqs.~\ref{eq:u_ev_DO} and \ref{eq:y_ev_DO}. However, the evolution equations written in the energetically ranked form are not differentiable with time whenever there is a singular value crossing. The issue of singular value crossing has been extensively studied in the context of using TDB for solving stochastic PDEs (\cite{CHZI13,CSK14,BCSK17,PB20}). This issue manifests itself through the appearance of the term $1/(\sigma_i - \sigma_j)$ in the evolution equations of TDB when written in the energetically ranked form. This can be understood intuitively: when the singular values cross, the more energetic mode before the crossing becomes the less energetic mode after the crossing. As a result, the more energetic mode, whose singular value is $\mbox{max} {\sigma_i,\sigma_j}$, varies non-differentiably with time. These singularities do not occur for the f-OTD evolution equations, and as a result, the f-OTD modes vary smoothly with time.

\subsubsection{Low-rank solution operator\label{sec:opt}}
In this section, we provide an operator interpretation of the f-OTD decomposition. We consider the MDE~\ref{eq:MDE_FOM} with a homogeneous initial condition, i.e., $\mathbf{V}(t=0) = \mathbf{0}$.  The low-rank operator for the MDE~\ref{eq:MDE_FOM}  is given by:
\begin{equation}
\mathbf{H}^t_{\mathcal{S}}(\cdot)  = \mathbf{\tilde {U}}( t)\bm{\Sigma}(t)\mathbf{\tilde {Y}}(t)^\mathrm{T} (\cdot ).
\end{equation}
 The input of the operator $\mathbf{H}^t_{\mathcal{S}}: \mathbb{R}^d  \rightarrow \mathbb{R}^{n}$ are the forcing coordinates $\mathbf{y} \in \mathbb{R}^d$ and the output of the operator is the state of the disturbance $\mathbf{v} \in \mathbb{R}^n$.  The above operator constructed by f-OTD is an approximation to the operator given by Eq.~\ref{eq:operator_full-rank}.  The above operator can be useful for a number of fluid dynamics problems. We highlight two of the key applications here:

\emph{(i) Surrogate modeling}:  The low-rank operator can serve as a rapid surrogate model to estimate responses to any forcing in $\mathcal{S}$. This is especially useful in receptivity analysis and uncertainty quantification. Once $\mathbf{H}^t_{\mathcal{S}}$ has been constructed, it allows for estimating responses to new forcings with minimal computational expense.

\emph{ (ii) Optimal forcing}:  Operator $\mathbf{H}^t_{\mathcal{S}}$ can be readily employed to find the most amplified forcing since it is factorized in the SVD form. The maximum disturbance energy is obtained by:
\begin{equation}
G(t) =\max_{ \mathbf{y} \neq \mathbf{0}} \dfrac{\|\mathbf{v}(t)\|^2_2}{\|\mathbf{y}\|^2_2} = \max_{ \mathbf{y} \neq \mathbf{0}} \dfrac{\|\mathbf{H}^t_{\mathcal{S}}(\mathbf{y})\|^2_2}{\|\mathbf{y}\|^2_2}  = \|\mathbf{\tilde{U}}( t)\bm{\Sigma}(t)\mathbf{\tilde{Y}}(t)^\mathrm{T}\|^2_2= \sigma^2_1(t).
\label{Optimal}
\end{equation}
Moreover, the first right and left singular vectors of $\mathbf{H}_{\mathcal{S}}^t$ are the optimal forcing  and the state of the optimal disturbance, respectively. Specifically, the coefficient of the optimal forcing that achieves the highest amplification at a given time $t^*$ is represented by the vector $\mathbf{y}^* = \mathbf{\tilde{y}}_1(t^*)$. The corresponding optimal forcing is denoted as $\mathbf{f}^* = \mathbf{F}\mathbf{\tilde{y}}_1(t^*)$, and the state of the optimal disturbance can be expressed as $\mathbf{v}^* = \sigma_1(t^*) \mathbf{\tilde{u}}_1(t^*)$.

%% file: DemonstrationCases.tex
\section{   Demonstration cases\label{Demonstration}}
In this section, we demonstrate the utility of f-OTD for various example problems.
\subsection{Toy model}

In the first example, we consider a three-dimensional nonlinear ordinary differential equation derived from the reduced-order modeling of laminar flow around a circular cylinder (\cite{noack2003hierarchy}). The toy model is represented by the following system of equations:
\begin{align*}
\dot{u}_1 &= \mu u_1 - \gamma u_2 - \alpha u_1u_3 - \beta u_1u_2, \\
\dot{u}_2 &= \gamma u_1 + \mu u_2 - \alpha u_2u_3 + \beta u_1^2, \\
\dot{u}_3 &= -\alpha u_3 + \alpha(u_1^2 + u_2^2).
\end{align*}
This reduced-order model corresponds to a periodically time-varying base flow and was recently analyzed using harmonic resolvent analysis (\cite{padovan2020analysis}). In our demonstration, we aim to show that a rank-1  f-OTD approximation (i.e., $r=1$) can accurately recover the most dominant direction of the forced linearized system instantaneously.
Performing linearization of the above reduced-order model  around the harmonically time-varying base state yields:
\[
\mathbf{L} = 
\begin{bmatrix}
    \mu-\alpha {u}_3-\beta {u}_2 && -\gamma-\beta{u}_1 && -\alpha{u}_1 \\
    \gamma+2\beta {u}_1 && \mu-\alpha{u}_3 && -\alpha{u}_2 \\
    2\alpha {u}_1 && 2\alpha {u}_2 && -\alpha
\end{bmatrix}\]
\\
\[
\mathbf{V} = 
\begin{bmatrix}
    \vline & \vline  \\
    \mathbf{v}_1 & \mathbf{v}_2 \\
    \vline & \vline 
\end{bmatrix},
\quad
  \mathbf{F} = 
\begin{bmatrix}
    \sin(\omega t) & 0 \\
    0 &  \cos(\omega t)  \\
    0 & 0 
\end{bmatrix}.
\]
The matrix $\mathbf{L} \in \mathbb{R}^{3\times 3}$, which represents the linearized system, is time-dependent due to its dependence on the current state of the system. The vector $\mathbf{v}_{i}$ is the response of the linearized system to the external forcing $\mathbf{f}_{i}(t)$, and 
  $\mathbf{F} =[\mathbf{f}_{1}(t), \mathbf{f}_{2}(t)] \in \mathbb{R}^{3\times 2}$  is the matrix of external forcing, consisting of periodic signals at the frequency $\omega$.  The chosen external forcing in our setup is similar to that employed in (\cite{padovan2020analysis}).

\begin{figure}
\centering
\subfigure[]{
\includegraphics[width=.45\textwidth]{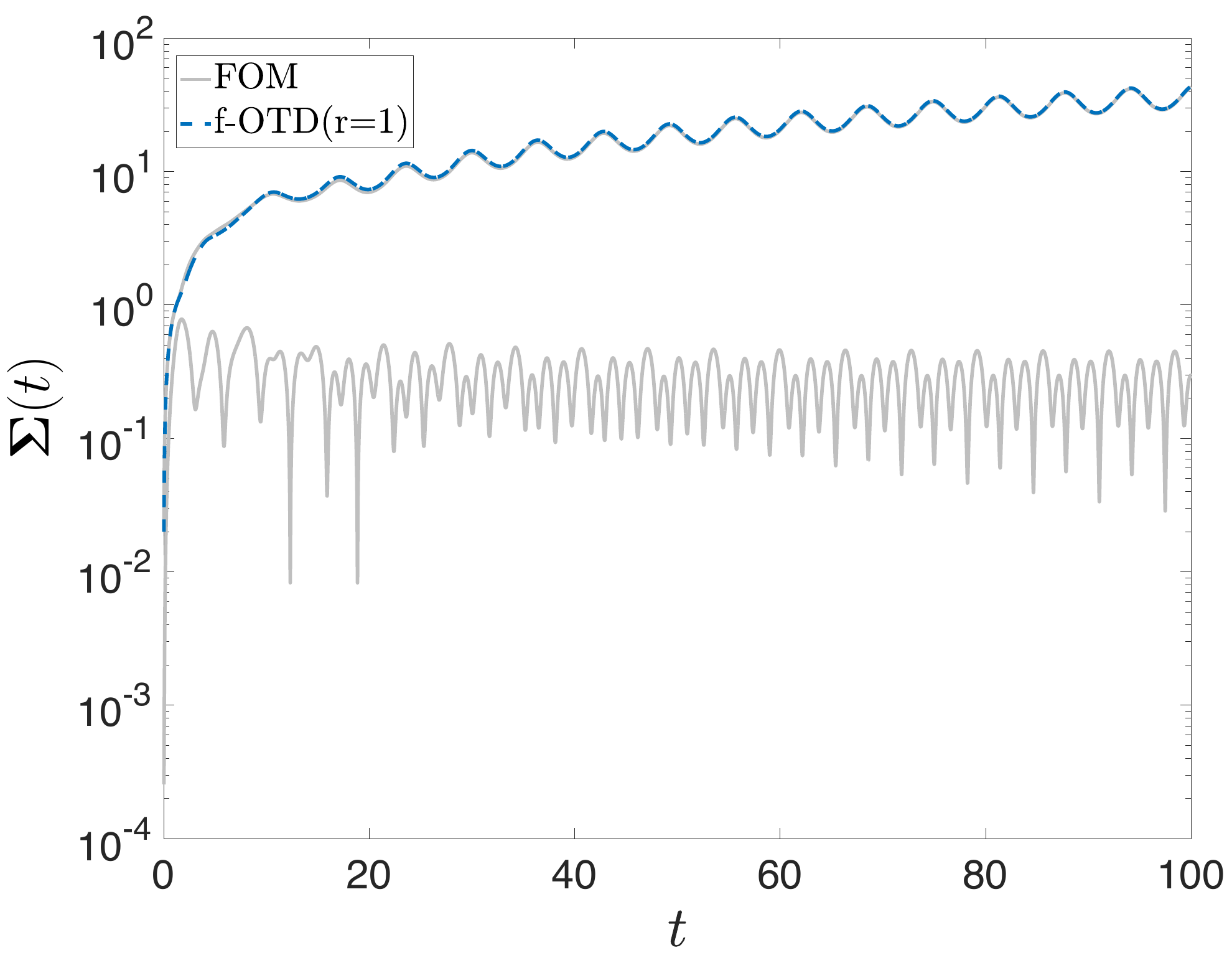}
\label{fig:ToyModel_SVD}
}
\subfigure[]{
\includegraphics[width=.45\textwidth]{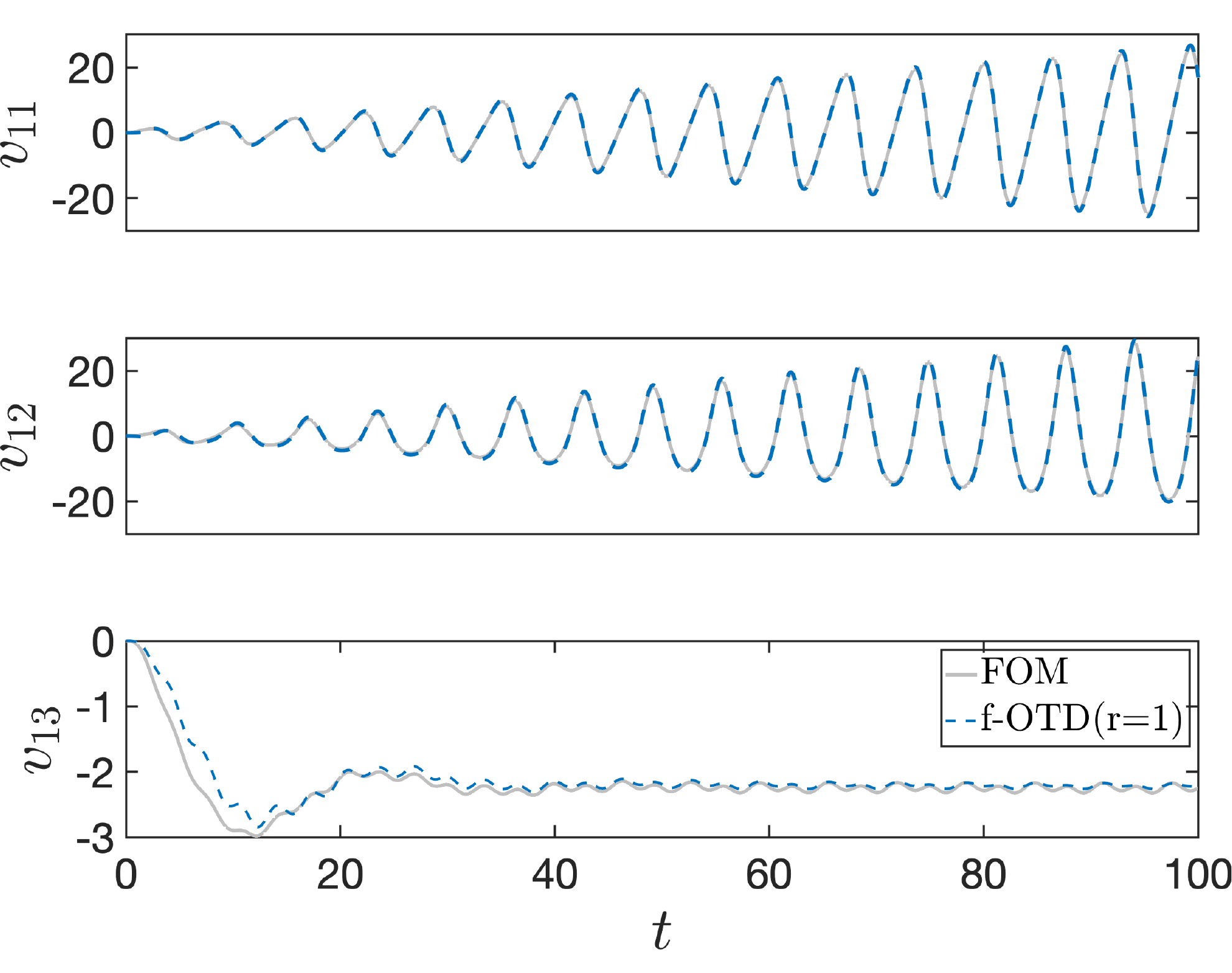}
\label{fig:ToyModel_Pert}
}
   \caption{Toy model: \text{(a)} Singular values obtained  from FOM and f-OTD for $r=1$. \text{(b)} Comparison of the reconstructed  first sensitivity vector $\mathbf{v}_1(t) =[v_{11}(t), v_{12}(t), v_{13}(t) ]^T$  obtained from  FOM and the rank-1 f-OTD   approximation.  }
  \label{fig:Toy}
\end{figure}

 To illustrate the utility of f-OTD, we choose a rank-$1$ f-OTD subspace ($r=1$) to approximate the response with a full-rank ($d=2$).  We integrate the nonlinear dynamical systems for $12 T$, where $T=2\pi/\omega$ and $\omega = 0.98$. Here, $T$ and $\omega$ are the limit cycle period and frequency of the nonlinear dynamical system, respectively.   Next, we solve the FOM with two different forcings for one time step to obtain $\mathbf{V}(t_0)$, where $t_0 = \Delta t = 0.01$. We then perform the SVD on $\mathbf{V}(t_0)$ to obtain the truncated rank-$1$ approximation of $\mathbf{V}(t_0)$, i.e., $\mathbf{V}(t_0) \approx \sigma_1(t_0) \mathbf{\tilde{u}}_1(t_0) \mathbf{\tilde {y}}_1(t_0)^\mathrm{T}$.  We then  initializes the rank-$1$ f-OTD subspace as follows: $\mathbf{u}_1(t_0) =\mathbf{\tilde{u}}_1(t_0)$ and $\mathbf{y}_1(t_0)=\sigma_1(t_0) \mathbf{\tilde{y}}_1(t_0)$.  For the time integration of f-OTD equations, we use the explicit fourth-order Runge–Kutta method (RK4). The ground truth is also obtained by solving FOM for $\mathbf{V}(t) = [\mathbf{v}_1(t), \mathbf{v}_2(t)]$. 

 The best time-dependent rank-$r$ approximation of the FOM solution is obtained by performing the SVD of  $\mathbf{V}(t)$ at each time step and truncating at rank $r$. Comparing the singular values derived from f-OTD with those from FOM serves as a valid metric for evaluating the performance of the f-OTD low-rank approximation. Figure \ref{fig:ToyModel_SVD} shows the temporal evolution of the singular values of the FOM and f-OTD.   The figure shows that  $\sigma_1(t)$ is two orders of magnitude greater than $\sigma_2(t)$, justifying approximating the disturbance matrix with a rank-$1$. Furthermore, we observe that the singular value obtained from f-OTD with rank-$1$ reduction agrees well with the most dominant singular value of the FOM.    The panels in Figure \ref{fig:ToyModel_Pert} show a comparison between the components of the first sensitivity vector $\mathbf{v}_1(t) =[v_{11}(t), v_{12}(t), v_{13}(t) ]^\mathrm{T}$ obtained from the f-OTD reconstruction and the FOM.  The f-OTD reconstruction of the first two components (${v}_{11}(t)$ and ${v}_{12}(t)$) agrees well with those obtained from the FOM --- in both the early transient growth stage as well the asymptotic saturation. However, a relatively larger error is observed for ${v}_{13}(t)$ in comparison to the other two components. This observation can be explained by recognizing that ${v}_{13}(t)$ is one order of magnitude smaller than ${v}_{11}(t)$ and ${v}_{12}(t)$, and for the optimal rank-$1$ approximation of the sensitivities, more precise representation of ${v}_{11}(t)$ and ${v}_{12}(t)$ is preferable to ${v}_{13}(t)$. This toy example demonstrates that f-OTD performs well when the base flow exhibits periodicity.

\subsection{Relationship between f-OTD and resolvent analysis}
The purpose of this example is to numerically demonstrate a connection between the f-OTD low-rank approximation and resolvent analysis.  To this end, we consider the linearized Burgers equation subject to harmonic forcing. We numerically demonstrate that the rank-$r$ f-OTD subspace asymptotically approximates the subspace obtained by the rank-$r$ SVD truncation of the resolvent operator. For this purpose, we consider the one-dimensional Burgers equation expressed as follows:
\begin{align}\label{eqn:ks_pde}
{\partial{ {u_b}}\over{\partial t}} + \frac{1}{2}{\partial{u^2_b}\over{\partial x}} = \nu{\partial^2{u_b}\over{\partial x^2}} , \quad x\in[0,L],
\end{align}
where the base flow is denoted by ${u_b}= {u_b}(x,t)$, $L=2\pi$ is the length of the domain, and the viscosity is given by $\nu=0.02$. We consider the initial condition ${u_b}(x,0)=\dfrac{1}{s \sqrt{2\pi}} \exp({-(x-\pi)^2/2 s ^2})$ with $s=0.15$.

Since our goal is to compare f-OTD with resolvent analysis, we linearized the Burgers equation around a time-invariant state. Specifically, we consider the time-averaged solution of Eq. \ref{eqn:ks_pde}:
\begin{equation}
    \overline{u}_b(x) = \frac{1}{T}\int_0^{T} u_b(x,t)dt.
\end{equation}
To investigate the amplification of disturbances around $\overline{u}_b(x)$, we consider the linear evolution equation for Burgers equation given by:
\begin{align}\label{eqn:ks_linearized_pde}
{\partial{ v_{i}}\over{\partial t}}= L(v_i)+ f_{i},
\end{align}
where $L(v_i) := -\dfrac{\partial (\overline{u}_b v_i)}{\partial x} + \nu \dfrac{\partial^2 v_i }{\partial x^2}$ denotes the linearized Burgers operator. The discretized representation of the linear operator is a time-invariant matrix. The harmonic  forcing is described by
\begin{equation}\label{eq:Burg_force}
f_{i}(x, t)=\mathbf{\delta}(x-x_i)\exp(j \omega t), \quad i=1,2, \dots, d,
\end{equation}
where ${x_i}$ represent the grid points. The forcing basis functions, $\mathbf{\delta}(x-x_i)$, are defined as $\mathbf{\delta}(x-x_i) =0 $ when $x \neq x_i$ and $\mathbf{\delta}(x-x_i) =1 $ when $x = x_i$. Here, $\omega =2\pi/T$ signifies the frequency with $T = 2$ being the period, and $d$ denotes the number of external excitations. We use \(d=n\) independent forcings, implying that the number of forcings equals the dimension of the system. The parameterization of the forcing as given by Eq.~\ref{eq:Burg_force} ensures that the f-OTD coefficient vectors ($\mathbf{y}_i \in \mathbb{R}^n$) are defined on the same grid as the state vectors.

For the spatial discretization of the Burgers equation, FOM, and f-OTD equations, we use the Fourier spectral method with \(n=256\) Fourier modes. For the temporal discretization, we use the exponential time-differencing fourth-order Runge-Kutta method (\cite{kassam2005fourth}) with a time step of \(\Delta t= 0.01\). We first solve the Burgers equation for one period ($T$). The mean state ($\overline{u}_b(x)$) is computed as the mean of $u_b(x,t)$ during this period. We solve the f-OTD equations using $\overline{u}_b(x)$ in the linearized operator. The f-OTD modes and coefficients are initialized by selecting the first \(r\) dominant modes from the FOM solution at the final time step of the period.

We now explain how f-OTD can be compared to resolvent analysis. While the f-OTD equations are solved in the temporal domain, the resolvent analysis is performed in the frequency domain. To facilitate this comparison, we convert the f-OTD reconstructed solution ($\mathbf{U}(t)\mathbf{Y}(t)^\mathrm{T}$) into the frequency domain using the Fast Fourier Transform (FFT). The FFT is applied to each column of the $\mathbf{U}(t)\mathbf{Y}(t)^\mathrm{T}$ matrix.

The f-OTD low-rank approximation is solved as an initial-value problem capturing both the transient evolution of the disturbance and its asymptotic behavior. As resolvent analysis captures the asymptotic behavior, we focus on comparing the f-OTD response after long-time integration. To this end, we solve the f-OTD evolution equations for 16 units of time. This duration ensures that the f-OTD reconstructed solution achieves a statically converged state.  Due to the absence of significant separation between the resolvent singular values, a relatively large f-OTD rank of $r=80$ is chosen. Therefore, $\mathbf{U}$ and $\mathbf{Y}$  are both of size $256 \times 80$ and   $\mathbf{H}^t_{\mathcal{S}} = \mathbf{U}(t)\mathbf{Y}(t)^\mathrm{T}$ is a rank-80 approximation of instantaneous solution operator. We then apply FFT to each column of  $\mathbf{H}^t_{\mathcal{S}}$ to obtain a frequency domain solution operator,  which is denoted as  $ \hat{\mathbf{H}}_{\mathcal{S}}(\omega)$. The temporal solution of the f-OTD asymptotically converged to a harmonic behavior with the same frequency as $\omega$. We use the f-OTD solution during the last two periods to compute $\hat{\mathbf{H}}_{\mathcal{S}}(\omega)$.

\begin{figure}
\centering
\subfigure[]{
\includegraphics[width=.45\textwidth]{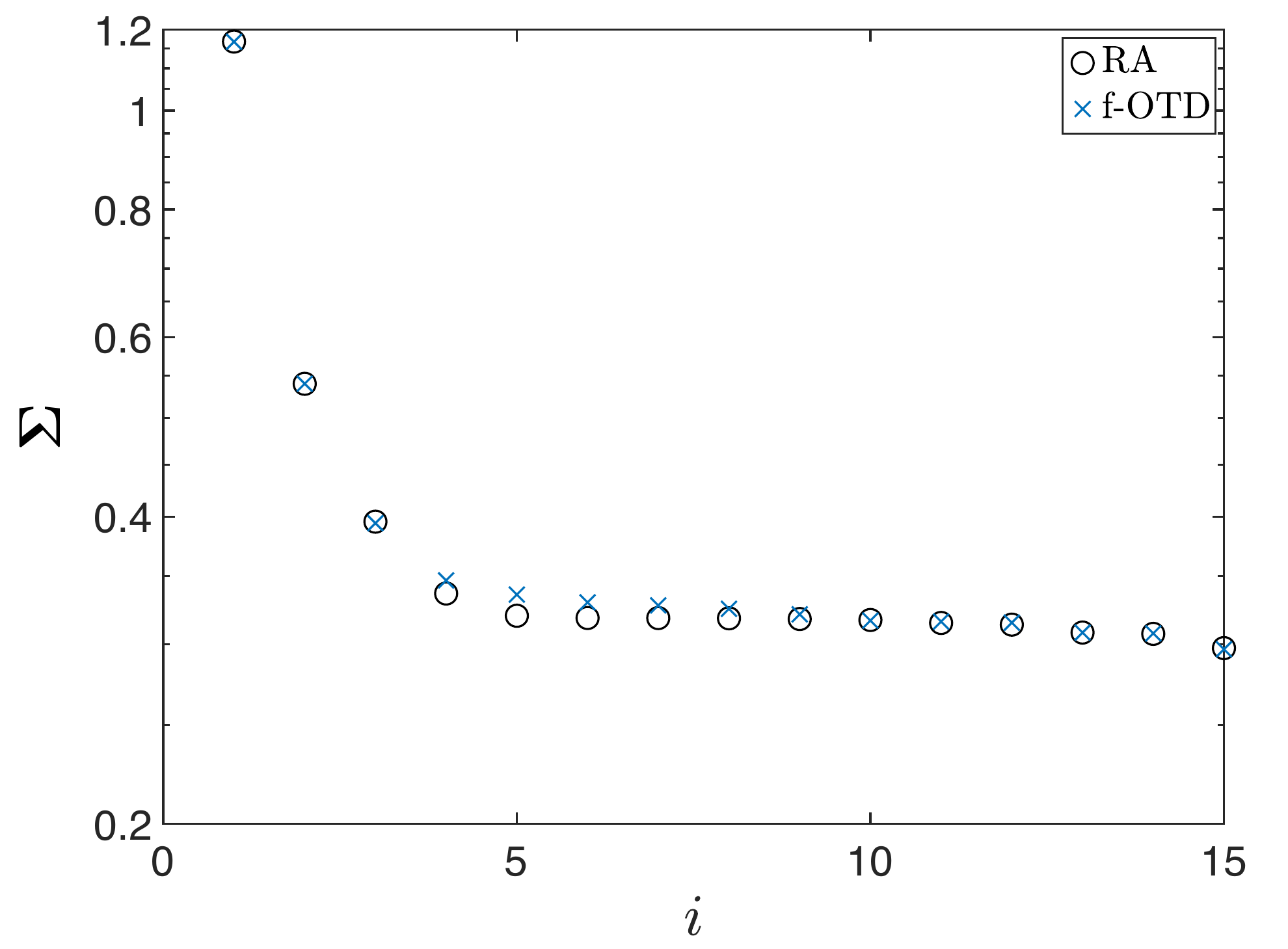}
\label{fig:Burgers_SVD_RA}
}
\subfigure[]{
\includegraphics[width=.45\textwidth]{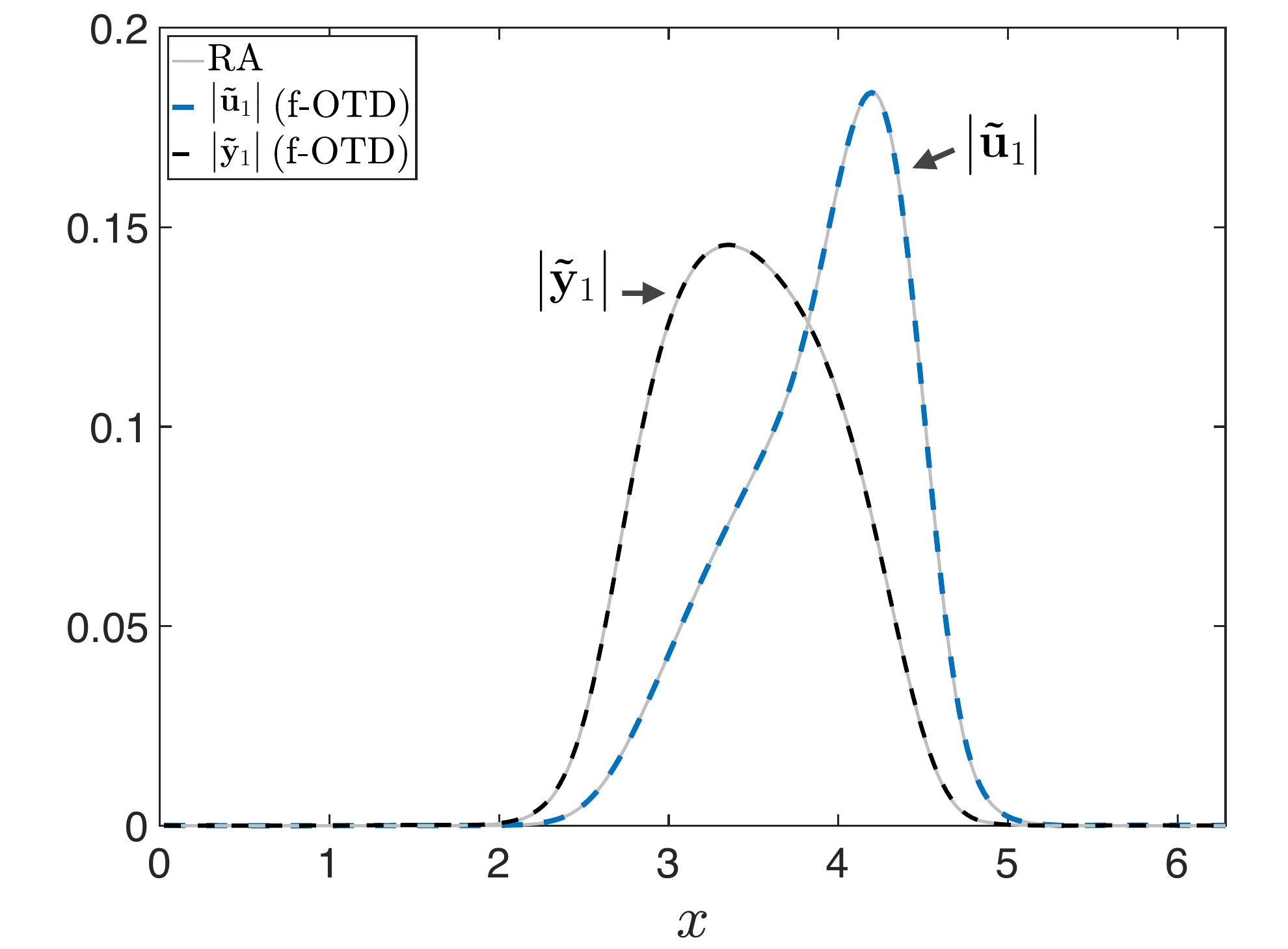}
\label{fig:Burgers_U_RA}
}
  \caption{Resolvent analysis (RA) and f-OTD for Burgers equation.  \text{(a)} Comparison of the first $15$ dominant singular values of the resolvent operator with the asymptotic solution of f-OTD. \text{(b)} The first dominant response and forcing mode from the resolvent operator and f-OTD.  }
  \label{fig:Burgurs_RA}
\end{figure}

The fifteen leading singular values of $\hat{\mathbf{H}}_{\mathcal{S}}(\omega)$  are compared against those obtained from resolvent analysis in  Figure \ref{fig:Burgers_SVD_RA}. The results indicate that the asymptotical solution of the f-OTD subspace and the resolvent analysis agree well with one another.   In Figure \ref{fig:Burgers_U_RA}, we compare the first dominant response and forcing modes obtained from resolvent analysis and the f-OTD approximation.   In particular,  the first left and right singular vectors of the resolvent operator and those obtained from f-OTD are shown. Excellent agreement between these singular vectors is observed. This example numerically demonstrates the connection between f-OTD and resolvent analysis. One might be able to rigorously establish a connection between f-OTD and resolvent analysis by analyzing the f-OTD response to a harmonically forced time-invariant linear system --- similar to the results that showed the asymptotic convergence of OTD modes ($\mathbf{F}=\mathbf{0}$) to the least unstable subspace of $\mathbf{L}$ \cite[Theorem 2.3]{babaee2016minimization}. However, such mathematical development is beyond the scope of the current study.

\subsection{ Temporally evolving jet} \label{2D_TEJ}
In this section, we demonstrate the capability of f-OTD for a transient flow with an arbitrary time-dependent base flow. 
 We consider the temporally mixing layer, which is a classic example of transient disturbance growth and has been extensively studied. See for example  (\cite{broze1994nonlinear, arratia2013transient}).  The transient evolution of the mixing layer exhibits time-varying dynamics, transient nonnormal growth, and vortex pairing, all of which make it a compelling test case for the f-OTD demonstration. A schematic representation of the mixing layer can be seen in Figure \ref{fig:Vorticity}. It is important to note that throughout the period under consideration ($0 \leq t \leq 30$), the base flow exhibits arbitrary time dependence. Additionally, the flow experiences a transition due to vortex pairing.

In the following, we use the f-OTD reconstructed input-output operator to find the worst-case disturbance, i.e., the most amplified disturbance. When the base flow is arbitrarily time-dependent, finding the most amplified disturbance requires solving an optimization problem, typically using a gradient-based optimization algorithm. Computing the gradient necessitates solving the adjoint equation, which must be done backward in time and requires storing the time-resolved nonlinear state. In general, adjoint equations are costly to solve for large-scale problems due to the input/output cost of storing and reading the nonlinear state from disk. Additionally, the iterations of gradient-based optimization algorithms have a sequential workflow due to the dependency of each iteration on the previous one, making them quite costly. 
 
\subsubsection{Problem setup} 

The evolution of the base flow is governed by the incompressible Navier-Stokes equations.  The f-OTD evolution equations are presented in Appendix \ref{Appendix_f-OTD_NS}.  Eqs. \ref{eq:NS_u}-\ref{eq:NS_v} are subject to periodic boundary conditions in $x$ and $y$ directions. The computational domain extends from \(x = 0\) to \(x = L\) and from \(y = 0\) to \(y = H\), with both \(L\) and \(H\) being equal to 1. For enhanced visualization, the contour plots are displayed over a domain with a length of \(2L\).
The initial velocity is given by $\mathbf u_b(x,y,t_0) = \bar {\mathbf u}(y) + \mathbf{u}'(x,y)$, where $\bar {\mathbf u}(y)$ is the initial velocity profile and $\mathbf{u}'(x,y)$ is divergence-free fluctuation velocity field to trigger the transition from laminar to an unsteady flow. The norm of  $\mathbf{u}'(x,y)$  is four orders smaller than the norm of \( \bar {\mathbf u}(y) \).  The velocity profile $\bar {\mathbf u}(y) = [\bar u_x(y), 0]^{\mathrm{T}}$ is defined as:
\begin{equation*}
\bar u_x(y) = \frac{u_{m}}{2}\left(\tanh\left(\frac{y-y_s}{\delta}\right) - \tanh\left(\frac{y-y_e}{\delta}\right) - 1 \right),
\end{equation*}
where $\delta = 0.01$, $y_s =  0.45$, $y_e = 0.55$, and $u_{m} = 1$ are used in this study. The jet is centered in the middle. We consider a Reynolds number given by \( Re = u_{m}H/\nu = 10^{4} \). 
 The forced linearized Navier-Stokes equations that govern the evolution of disturbances due to external forcing as well as the f-OTD equations for the incompressible Navier-Stokes equations are given in Appendix \ref{Appendix_f-OTD_NS}.
Eqs. \ref{eq:LNS_u}-\ref{eq:LNS_div} constitute the FOM in this demonstration.

\begin{figure}
  \centering
    \includegraphics[width=.99\linewidth]{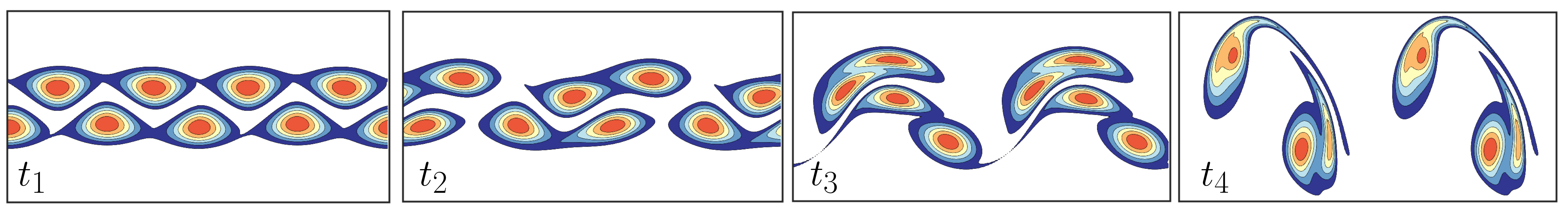}
  \caption{Schematic of the temporally evolving jet. The evolution of the base flow at different time instances ($t_1=12$, $t_2 =15$, $t_3 =18$, $t_4=21$). }
  \label{fig:Vorticity}
\end{figure}

The two-dimensional incompressible Navier-Stokes equations, FOM, and the evolution equations for f-OTD are solved using the Fourier spectral method, employing \(N_x = N_y = 2^7\) Fourier modes in each direction. For the time integration, we utilize a fourth-order Runge-Kutta scheme with a time step of \(\Delta t = 3.125 \times 10^{-3}\). To initialize the time-varying base flow, we integrate the incompressible Navier-Stokes equations for 25 time units. Subsequently, the solution from the final time step is taken as the initial condition for the base flow. Figure \ref{fig:Vorticity} depicts the base flow evolution at various time instances.


 \subsubsection{Harmonic forcing} 

 To generate the localized forcing, we introduce the following expressions:
\begin{align}
\tilde{f}_{x}(x,y)&= \frac{I(y)}{k_x } \cos(2\pi k_x x)\sin(2\pi k_y y), \label{eq:localized_forcing_x}\\
\tilde{f}_{y}(x,y)&= -\dfrac{I(y)}{k_y }  \sin(2\pi k_x x)\cos(2\pi k_y y),
\label{eq:localized_forcing_y}
\end{align}
where $I(y) = \tanh\Big(\dfrac{(y-y_s)}{\delta}\Big) - \tanh\Big(\dfrac{(y-y_e)}{\delta}\Big)$ serves as the indicator function that localizes the forcing within the region defined by $y_s \leq y \leq y_e$ and different forcings are generated by varying the wavenumbers $k_x$ and $k_y$. 
The vector field $\tilde{\mathbf f}= [\tilde{f}_{x},\tilde{f}_{y}]^{\mathrm{T}}$ is not divergence-free due to the multiplication of $I(y)$ by each component. We map this vector field to a divergence-free vector field using  the projection function $\phi$, which is obtained by solving $\nabla^2 \phi = \nabla \cdot \tilde{\mathbf f}$ and updating the vector field using the following relations:
\begin{align}
f_{x}(x,y,t)&=c (\tilde{f}_x(x,y) - \frac{\partial \phi}{\partial x})\sin(\omega t), \label{eq:forcing_x}\\
f_{y}(x,y,t)&=c (\tilde{f}_y(x,y) - \frac{\partial \phi}{\partial y})\sin(\omega t).
\label{eq:forcing_y}
\end{align}


It is easy to verify that the vector field $(f_x,f_y)$ is divergence-free. Here, the coefficient $c$ determines the energy spectrum of each force component and is given by $c=1/(k_x+k_y)$.  We consider a total of $d=144$ external excitations by varying the wave numbers $k_x=[1,2, \dots, 12]$ and $k_y=[1,2, \dots, 12]$. We consider the tensor product of these sets and generate $d=144$ forcings.  The temporal forcing frequency is denoted by \( \omega \), and for all the forcings considered in this problem, we set \( \omega = 0.37 \). This excitation frequency is identified as one of the dominant frequencies in the nonlinear base flow.


The space-time discretization of the f-OTD equation is identical to that of the FOM.
To initialize the f-OTD modes and modal coefficients, we solve the FOM for one time step for all \( d=144 \) forcings. We construct a matrix of the FOM solution, wherein the matrix has \( 2N_x N_y = 2^{15} \) rows, corresponding to the total number of grid points times two for the two components of velocity, and \( d=144 \) columns, with each column representing the response to each forcing. We then perform SVD on this matrix and set the rank-$r$ matrix SVD-truncated matrix as the initial condition for the f-OTD subspace. For very large dynamical systems, the initial condition for the f-OTD matrices can be computed through a targeted sparse sampling of the FOM to reduce computational costs (\cite{DPNFB23}.


  \begin{figure}
  \centering
    \includegraphics[width=.55\linewidth]{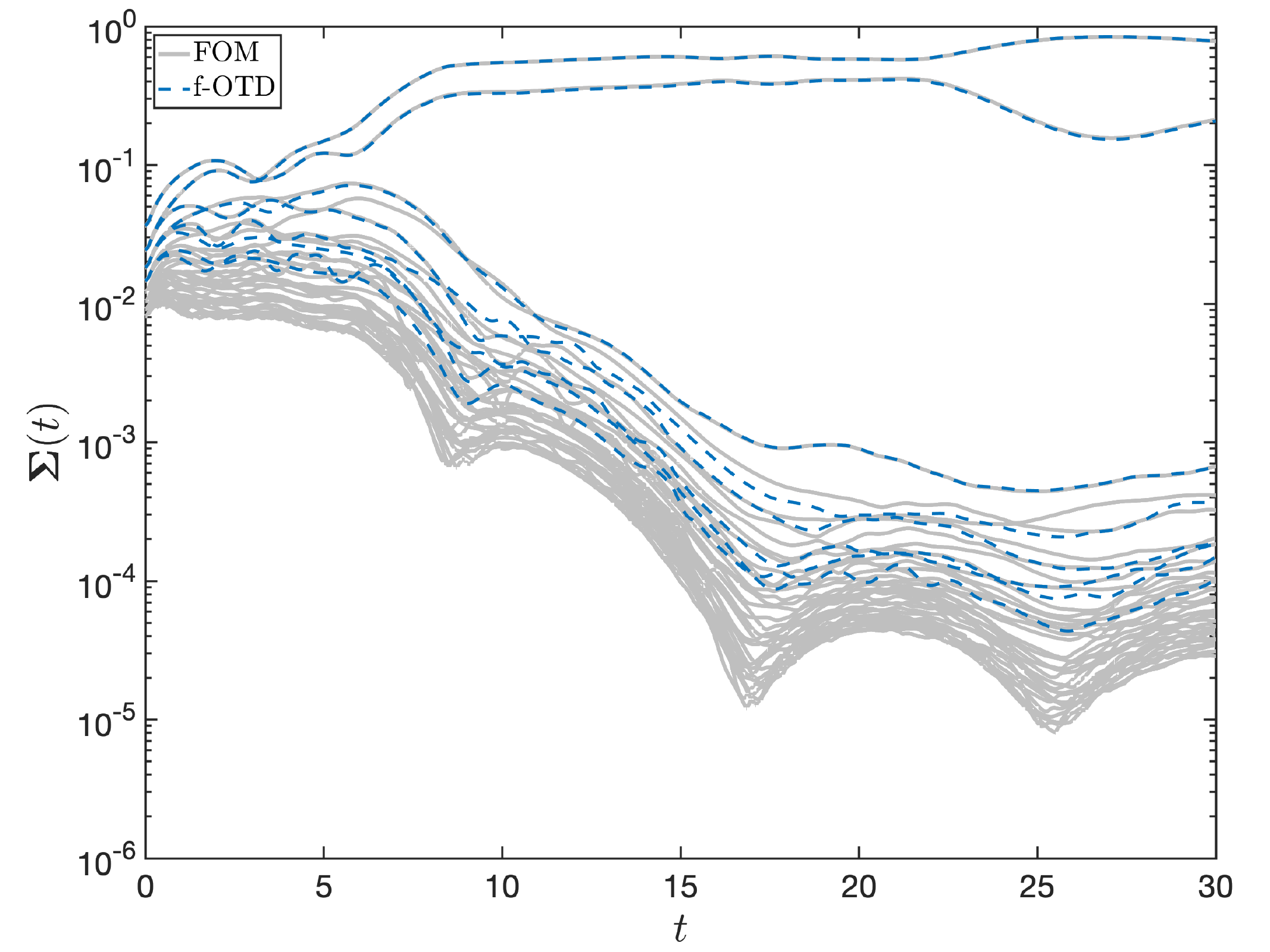}
   \caption{Temporally evolving jet: comparison of the dominant normalized singular values obtained from FOM and f-OTD.  } 
  \label{fig:SVD_Temporally}
\end{figure}

Figure \ref{fig:SVD_Temporally} displays a comparison of normalized singular values between FOM and f-OTD. To compute the normalized singular values, we divide the instantaneous singular value from f-OTD and FOM by the total sum of the FOM singular values at each time instance.  We solve the f-OTD equations with a rank of \( r=8 \), and Figure \ref{fig:SVD_Temporally} illustrates that the largest singular values are accurately approximated by f-OTD.
The dimensionality of this system varies over time, as observed by the clustering of singular values at the beginning (\( 0 \leq t \leq 5 \)) where many modes are required to approximate the FOM. As the flow evolves, two time-dependent f-OTD modes become dominant—growing by two to three orders of magnitude larger than the other modes. The dominance of the first two singular values suggests that the response to various forcings can be accurately approximated with a low-rank approximation based on time-dependent bases.


\begin{figure}
\centering
\includegraphics[width=0.72\textwidth]{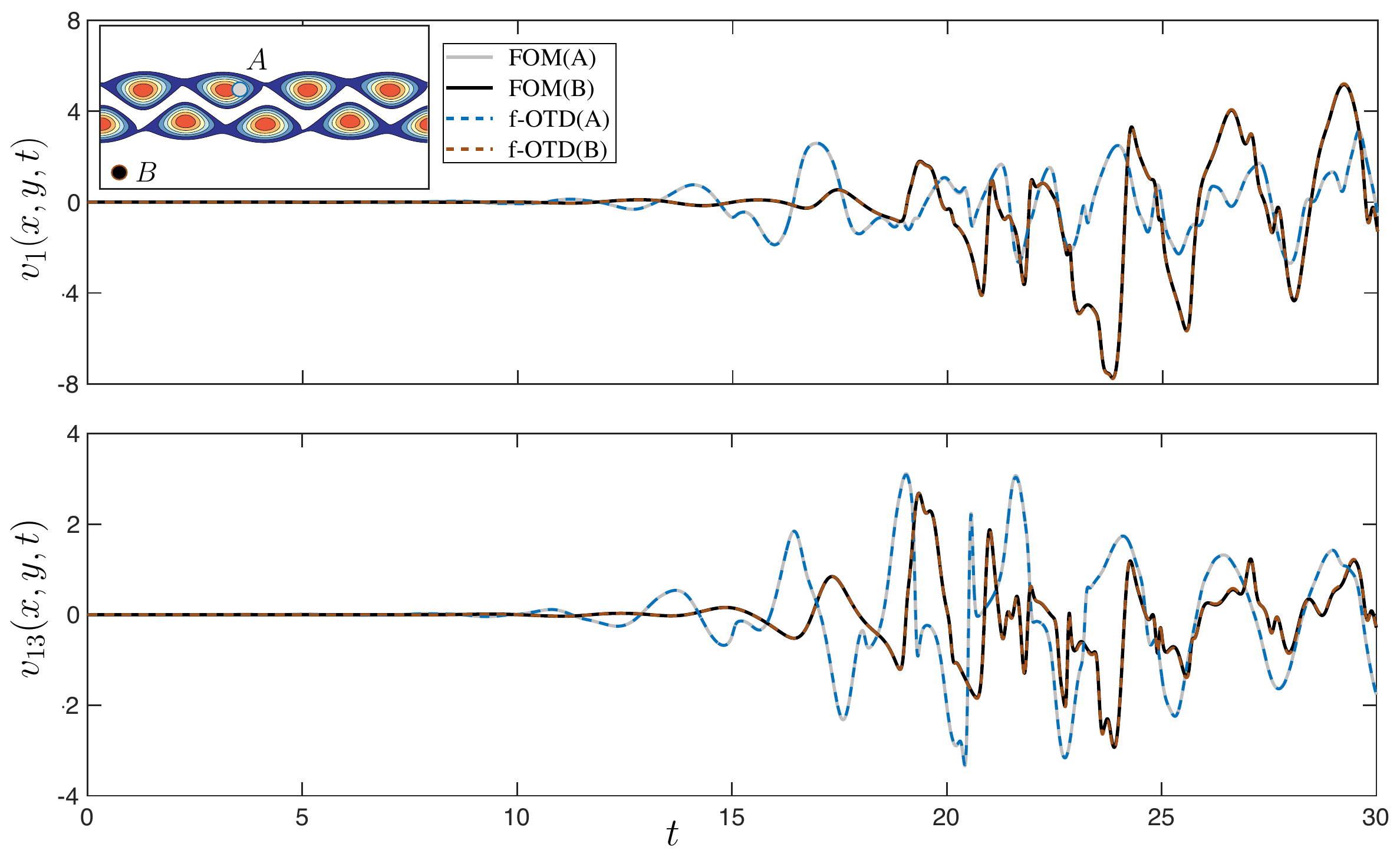}
\caption{Temporally evolving jet:  comparison of the temporal evolution of the perturbation velocity at two probe locations of $(x_A,y_A) = (0.39, 0.59)$ and $(x_B,y_B) = (0.10, 0.12)$.}
\label{fig:Reconstruction}
\end{figure}

Figure \ref{fig:Reconstruction} demonstrates the efficacy of f-OTD reduction in reconstructing the disturbance field when there is a strong cross-frequency interaction between the forcing frequency and the base flow. We examine two distinct excitations, \(v_1\) and \(v_{13}\), corresponding to wavenumbers \( (k_x, k_y) = (1, 1) \) and \( (k_x, k_y) = (2, 1) \), respectively. The figure presents a comparison of the temporal evolution of the f-OTD reconstructed response and the FOM at two different locations, \( A \) and \( B \), with coordinates \( (x_A, y_A) = (0.39, 0.59) \) and \( (x_B, y_B) = (0.10, 0.12) \). The disturbance exhibits transitional multi-frequency behavior, attributable to the cross-frequency interactions between the harmonic forcing frequency and the arbitrarily time-dependent base flow. It is observed that f-OTD low-rank approximation captures this evolution accurately.


\begin{figure}
\centering
\includegraphics[width=1\textwidth]{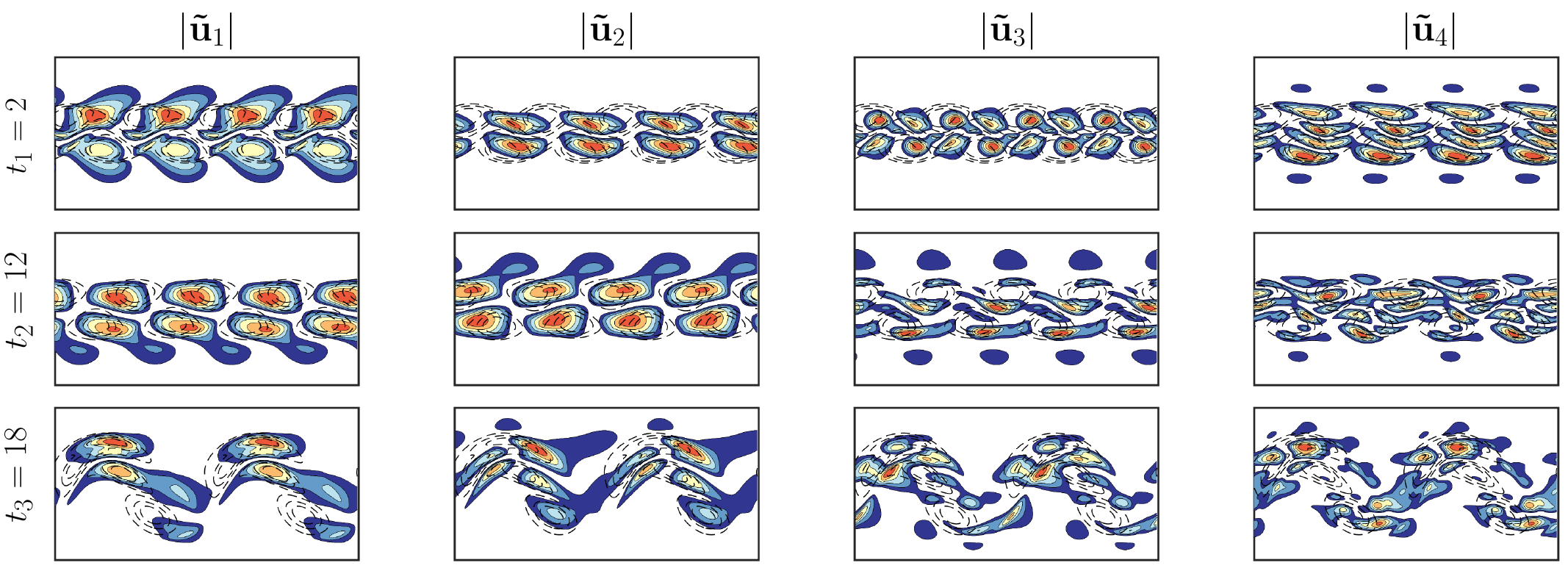}
\caption{Temporally evolving jet: the first four dominant left singular vectors of the f-OTD low-rank solution operator at different times (dashed line shows the base flow's vorticity). The f-OTD modes are energetically ranked in descending order.}
  \label{fig:f-OTD}
\end{figure}

Figure \ref{fig:f-OTD} displays snapshots of the first four dominant f-OTD spatial modes at different time instances, sorted by their singular values from the most dominant to the least dominant modes. The dashed lines show the contours of vorticity for the base flow.   The first two modes are associated with the dominant singular values, $\sigma_1(t)$ and $\sigma_2(t)$, and exhibit similar structures. The dominant modes capture the largest structures, while the lower modes extract finer structures. 
It is also evident that the f-OTD modes evolve instantaneously with the base flow.


\begin{figure}
\centering
\includegraphics[width=1\textwidth]{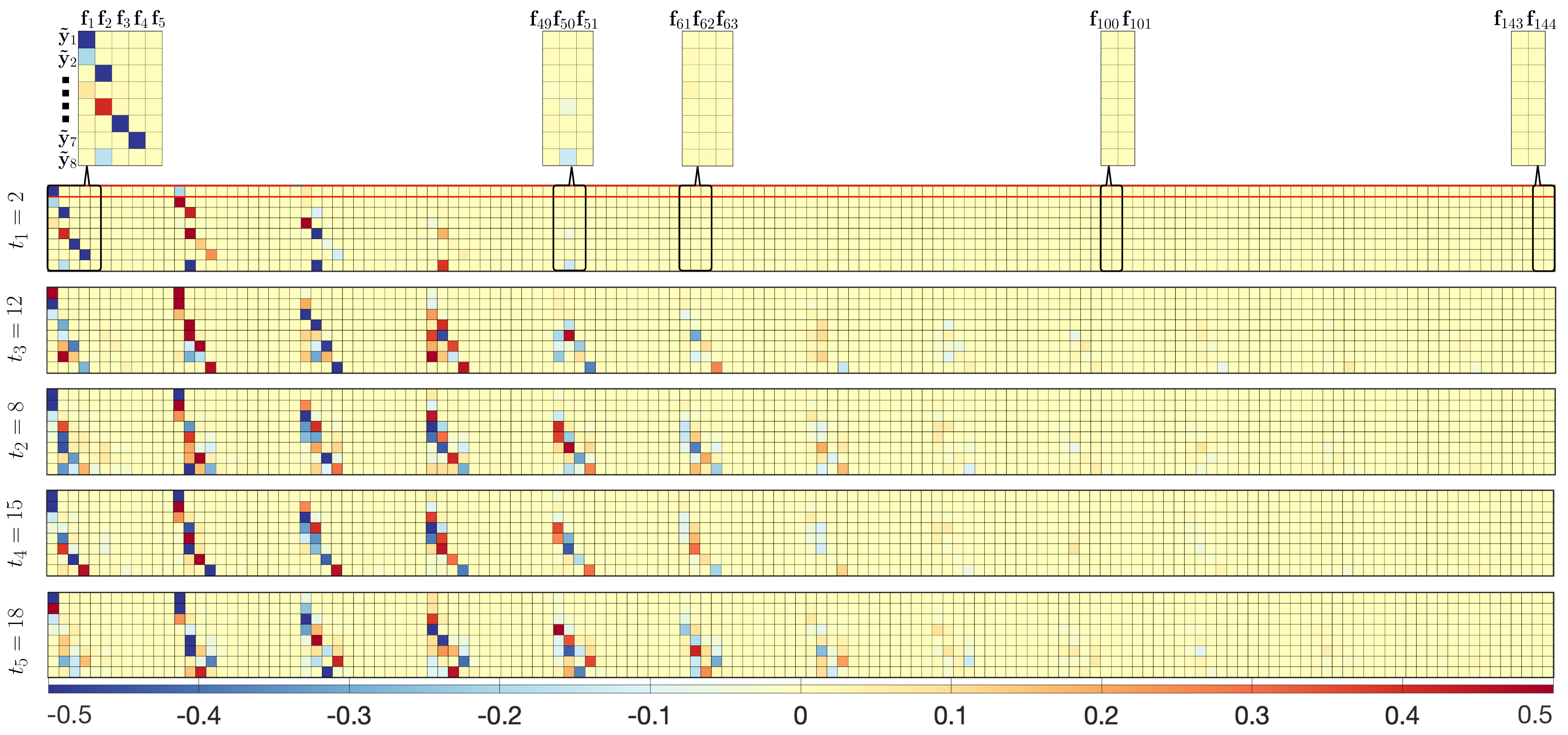}
\caption{Temporally evolving jet:   the first eight dominant right singular vectors ($\mathbf{\tilde Y}^\mathrm{T}(t)$) of the f-OTD low-rank solution operator at different times.}
  \label{fig:Y_f-OTD}
\end{figure}

Figure \ref{fig:Y_f-OTD} shows the time evolution of the orthonormalized modal coefficients $\mathbf{\tilde Y}(t)^\mathrm{T}$. Each row approximates a right singular vector of the solution operator, while each column represents the contribution of each excitation to different modes. As time progresses, the modal coefficients evolve, capturing the time-dependent contribution of different forces to the f-OTD low-rank subspace. For example, at $t_1=2$, the first row ($\mathbf{\tilde y}_1^\mathrm{T}$) shows that among all $144$ excitations, $\mathbf{f}_1$ and $\mathbf{f}_{13}$, associated with wave numbers $(k_x,k_y)=(1,1)$ and $(2,1)$ respectively, have the most significant contributions. However, the third row ($\mathbf{\tilde y}_3^\mathrm{T}$), associated with $\sigma_3$, demonstrates the dominant contributions of $\mathbf{f}_2$ and $\mathbf{f}_{14}$. The modal coefficients contain rich insights into the excitations and can be exploited for physical or dynamical understanding.

\begin{figure}
\centering
\includegraphics[width=1\textwidth]{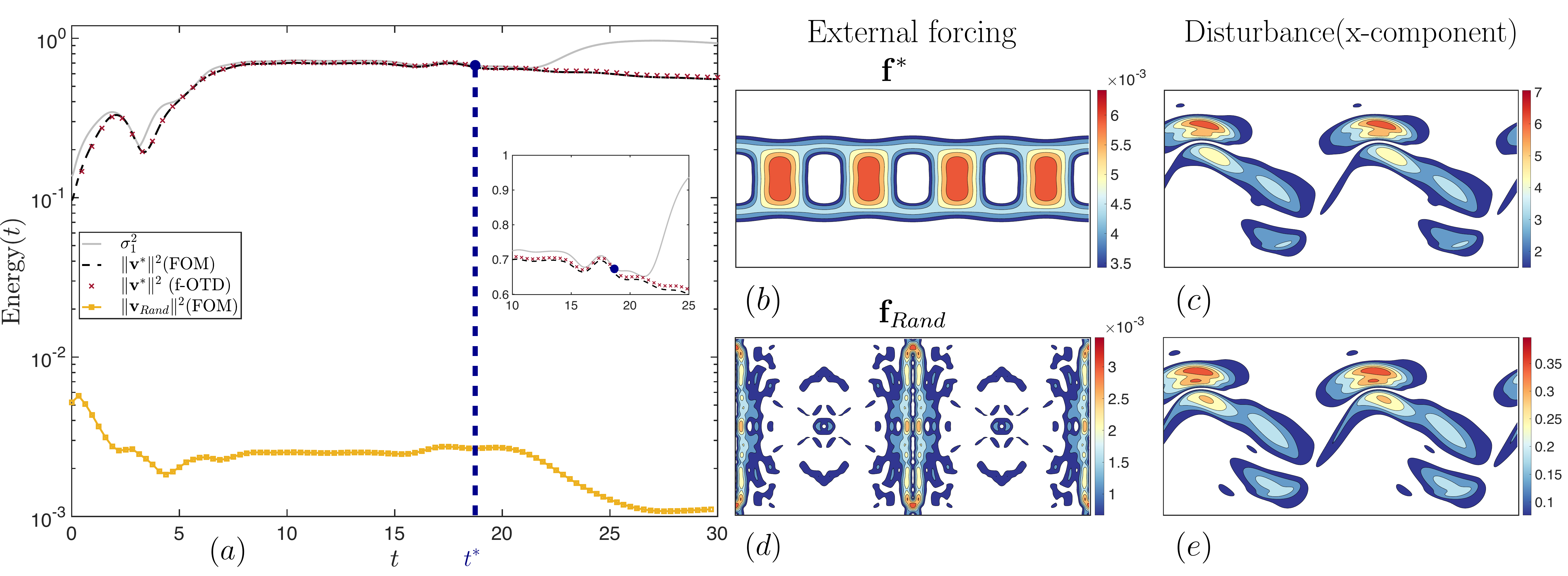}
\caption{Temporally evolving jet: comparison between optimal forcing and a random forcing within the forcing space.  At time $t^*=18.75$, the magnitude of the optimal and random forcing with their corresponding responses are displayed.}
\label{fig:Optimum_forcing_response}
\end{figure}

We utilize the low-rank approximation of the solution operator based on f-OTD to determine the optimal forcing without incurring additional computational costs. In particular, we seek to identify the forcing $\mathbf{f} \in \mathcal{S}$ whose response achieves maximum amplification at $t = t^*$ among all forcings in $\mathcal{S}$. As demonstrated in Section \ref{sec:opt}, the optimal forcing is obtained from $\mathbf{f}^* = \mathbf{F}\mathbf{\tilde y}_1(t^*)$, leading to maximal amplification at the designated time $t^*$. To obtain the optimal perturbation vector, we solved Eq.~\ref{eq:state_2} with the forcing $\mathbf{f}=\mathbf{F}\mathbf{\tilde y}_1(t^*)$.  
To underscore the significance of the optimal  forcing vector $\mathbf{\tilde y}_1(t^*)$, we solved Eq.~\ref{eq:state_2} for a random excitation, $\mathbf{f}_{\text{Rand}} = \mathbf{F}\mathbf{y}_{\text{Rand}}$, where $\|\mathbf{y}_{\text{Rand}}\|$=1. In Panel \ref{fig:Optimum_forcing_response}$\text{a}$, we compare the evolution of disturbance energy resulting from these two different external forcings. The gray line represents the maximum possible amplification $G_{\text{max}}(t)=\sigma_1^2(t)$, which signifies the maximal energy that any $\mathbf{f} \in \mathcal{S}$ can achieve; it can be regarded as the envelope of perturbation energy for all $\mathbf{f} = \mathbf{F}\mathbf{y}$ where $\|\mathbf{y}\|=1$.
 It is evident that the response to the optimal forcing, shown by the dashed black line, reaches the envelope ($\sigma^2_{1}$) at $t^*$, shown by a blue circle, after which its energy decays. In contrast, the energy of the random excitation experiences growth but remains roughly one to two orders of magnitude smaller than the maximum possible amplification.

 To demonstrate the capability of f-OTD as a rapid surrogate model, we obtain state disturbance due to  the optimal forcing using the operator:
 \begin{equation*}
\mathbf{v}^*(t)=\mathbf{H}^t_{\mathcal{S}}(\mathbf{\tilde y}_1(t^*))  = \mathbf{\tilde{U}}( t)\bm{\Sigma}(t)\mathbf{\tilde{Y}}(t)^\mathrm{T} (\mathbf{\tilde y}_1(t^*) ).
\end{equation*} 
Note that $\mathbf{\tilde y}_1(t^*)$ is not a time-dependent vector. The energy of this solution is shown with cross-marker symbols, which matches well with the disturbance obtained by solving  Eq.~\ref{eq:state_2}, which is the ground truth in this setting.


Panels \ref{fig:Optimum_forcing_response}$\text{b-e}$ illustrate the structures of these excitations and their corresponding responses at time $t^*$.  The $x$-component of the response to the forcing is shown in panels (\ref{fig:Optimum_forcing_response}$\text{c}$ and 
 \ref{fig:Optimum_forcing_response}$\text{e}$). 
The optimal forcing in Panel \ref{fig:Optimum_forcing_response}$\text{b}$ exhibits a significant presence in the shear layer and is influenced by lower wave numbers. In panels \ref{fig:Optimum_forcing_response}$\text{c}$ and \ref{fig:Optimum_forcing_response}$\text{e}$ we observe that the flow responses share similar structures, however, they have completely different magnitudes. The fact that the disturbance resulting from the random forcing has a shape similar to that of the optimal disturbance is because  $\mathbf{y}_{\text{Rand}}$ has a small projection onto the optimal $\mathbf{\tilde y}_1(t^*)$. Given that this direction in the forcing space grows significantly faster than other directions, its response ultimately dominates the shape of the disturbance.


\subsection{Two-dimensional chaotic Kolmogorov flow}\label{2D_Kolmog}

The purpose of this demonstration is to show the performance of f-OTD in computing the response of a chaotic flow subject to high-dimensional external excitations.   In particular, we consider a chaotic flow that exhibits a short-lived intermittent phenomenon.

We consider the two-dimensional chaotic Kolmogorov flow (\cite{PSF91}) governed by the forced incompressible Navier-Stokes equations as shown below
\begin{equation}
\frac{\partial \mathbf u_b}{\partial t} + (\mathbf u_b \cdot \nabla) \mathbf u_b = - \nabla p_b + \frac{1}{Re}\nabla^2 \mathbf u_b + \mathbf f_b, \quad \quad  \nabla \cdot \mathbf u_b = 0,
\end{equation}
where  $\mathbf u_b(x,y,t) =[u_b(x,y,t),v_b(x,y,t)]^\mathrm{T}$ is the time-dependent base flow and $\mathbf f_b(y) = \sin (n y) \mathbf e_1$, where $\mathbf e_1 = [1, 0]^\mathrm{T}$. The flow is considered in the physical domain of $(x,y) \in [0, 2\pi]\times [0, 2\pi]$ with periodic boundary conditions in both $x$ and $y$. The Kolmogorov flow has a laminar solution
\begin{equation}\label{eq:base_SS}
    \mathbf u_b(y) = \frac{Re}{n^2} \sin (n y) \mathbf e_1.
\end{equation}
The above laminar profile is stable for $n=1$ and any Reynolds number. For $n>1$ and a sufficiently large Reynolds number, the above laminar solution is unstable (\cite{PSF91}) . 

The Kolmogorov flow has applications in magnetohydrodynamics, where it can be reproduced using appropriate magnetic and electric fields. For a recent review of the Kolmogorov flow and its applications, see \cite{F18}.
 
We consider $n=4$ and $Re=40$, which result in an unstable flow. Under these parameters, the flow is chaotic and exhibits a strange attractor. Moreover, for these parameters, the Kolmogorov flow exhibits intermittent dynamics, which manifests itself as short-lived burst of energy dissipation. The energy dissipation is defined as:
\begin{equation}
    D(t) = \frac{1}{ReL^2 }\int_0^{L} \int_0^{L} |\omega(x,y,t) |^2 dx dy,
\end{equation}
where $\omega(x,y,t)$ is the vorticity and $L=2\pi$ is the length of the domain in each dimension. Also, the energy input is defined as: 
\begin{equation}
    I(t) = \frac{1}{L^2 }\int_0^{L} \int_0^{L} \mathbf u_b(x,y,t)\cdot \mathbf f_b(y) dx dy.
\end{equation}

\begin{figure}
\centering
\includegraphics[width=0.99\textwidth]{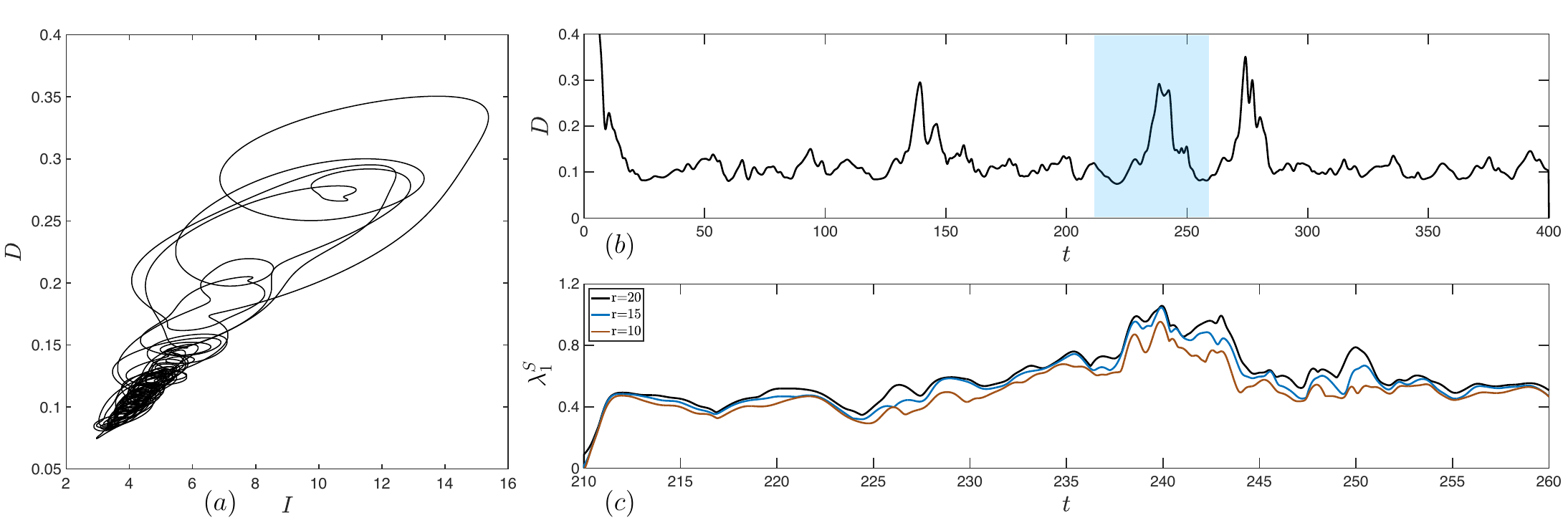}
\caption{The Kolmogorov flow: (a) energy input versus energy dissipation; (b) evolution of the energy dissipation $D$;  and (c)  largest eigenvalue, $\lambda^S_1$, of the symmetric part for reduced linear matrix  $\mathbf S_r$ for $d=16,384$.}
\label{fig:Lr_diss}
\end{figure}

In Figure \ref{fig:Lr_diss}a, $D(t)$ versus $I(t)$ is shown.  This figure portrays that chaotic attractor as well as the excursion from the attractor.

The bursting phenomenon was studied by \cite{FS16}, where the OTD low-rank approximation is used to build indicators for these extreme events. We also consider the same parameters as \cite{FS16}, i.e., $n=4$ and $Re=40$. In this work, we investigate the effect of external forcing disturbance as opposed to the initial condition perturbations considered by \cite{FS16}.

We use f-OTD for the numerical simulation of the Kolmogorov flow subject to high-dimensional external disturbances. In particular, we consider divergence-free forcing disturbances in the form of:
\begin{equation*}
f_{x}(x,y)= \frac{c}{k_x }\cos(k_x x) \sin(k_y y)   \quad \quad 
f_{y}(x,y)= - \dfrac{c}{k_y} \sin(k_x x)\cos(k_y y),
\end{equation*}
where $c =1/(k_x^2+k_y^2)$ is the forcing spectrum. Here,  $k_x$ and $k_y$  are the wave numbers ranging from $1$ to $p$, i.e., $k_x=[1,2\dots, p]$ and $k_y=[1,2\dots, p]$. Therefore, the total number of external excitations considered is $d=p^2$.  The external forcing defined above should be viewed as infinitesimal disturbances to the base flow as shown below:
\begin{equation}
\frac{\partial \mathbf u'_b}{\partial t} + (\mathbf u'_b \cdot \nabla) \mathbf u'_b = - \nabla p'_b + \frac{1}{Re}\nabla^2 \mathbf u'_b + \mathbf f_b + \epsilon \mathbf f,
\end{equation}
where $\mathbf f = [f_x, f_y]^\mathrm{T}$ is the external disturbance vector and $\mathbf u'_b$ and $p'_b$ are the perturbed velocity  and pressure fields, respectively.  Therefore, the disturbance field is $\mathbf v = (\mathbf u'_b - \mathbf u_b)/\epsilon$ as $\epsilon \rightarrow 0$. The evolution of $\mathbf v$ is governed by the forced linearized Navier-Stokes Eqs. \ref{eq:LNS_u} and \ref{eq:LNS_div} presented in  Appendix A.  Different forcings $\mathbf f_i$, for $i=1,2,\dots, d$ are obtained by considering all combinations of wavenumbers $k_x$ and $k_y$ in their respective ranges, which generates the disturbance $\mathbf v_i$.   The base flow driving force ($\mathbf{f_b}$) does not appear in the linearized Navier-Stokes equations because $\mathbf{f_b}$ is not a function of the state variables and, therefore, is unperturbed.

We solve the base flow and the f-OTD equations using the same numerical methods as in the previous example. We consider a grid of size $128 \times 128$ and $\Delta t = 0.004$. 

We consider two cases: $p=10$ and $p=128$, which results in $d=10^2 = 100$ and $d=128^2=16,384$, respectively. The dimension of the space spanned by the forcing in the case of $d=128^2$ is equal to the size of the grid points.  The case with $p=10$ is considered for verification purposes, for which we solve the FOM. For the case of $p=128$, solving the FOM is too costly because it requires solving $16,384$ linearized Navier-Stokes equations. For this case, we only perform the f-OTD simulations.

We first integrate the base flow until it reaches the attractor. To this end, we initialize the base flow with the steady-state profile given by Eq. \ref{eq:base_SS}, superimposed with small divergence-free perturbations. For the chosen parameters, the base flow is unstable and becomes chaotic and arbitrarily time-dependent. We integrate the flow until $t_0=210$ after which an intermittent phenomenon occurs as it is evident in the dissipation shown in Figure \ref{fig:Lr_diss}b.  We then solve the FOM for one timestep and then compute the SVD of the FOM and truncate at rank $r$ to initialize the f-OTD low-rank matrices.  

\begin{figure}
\centering
\includegraphics[width=0.8\textwidth]{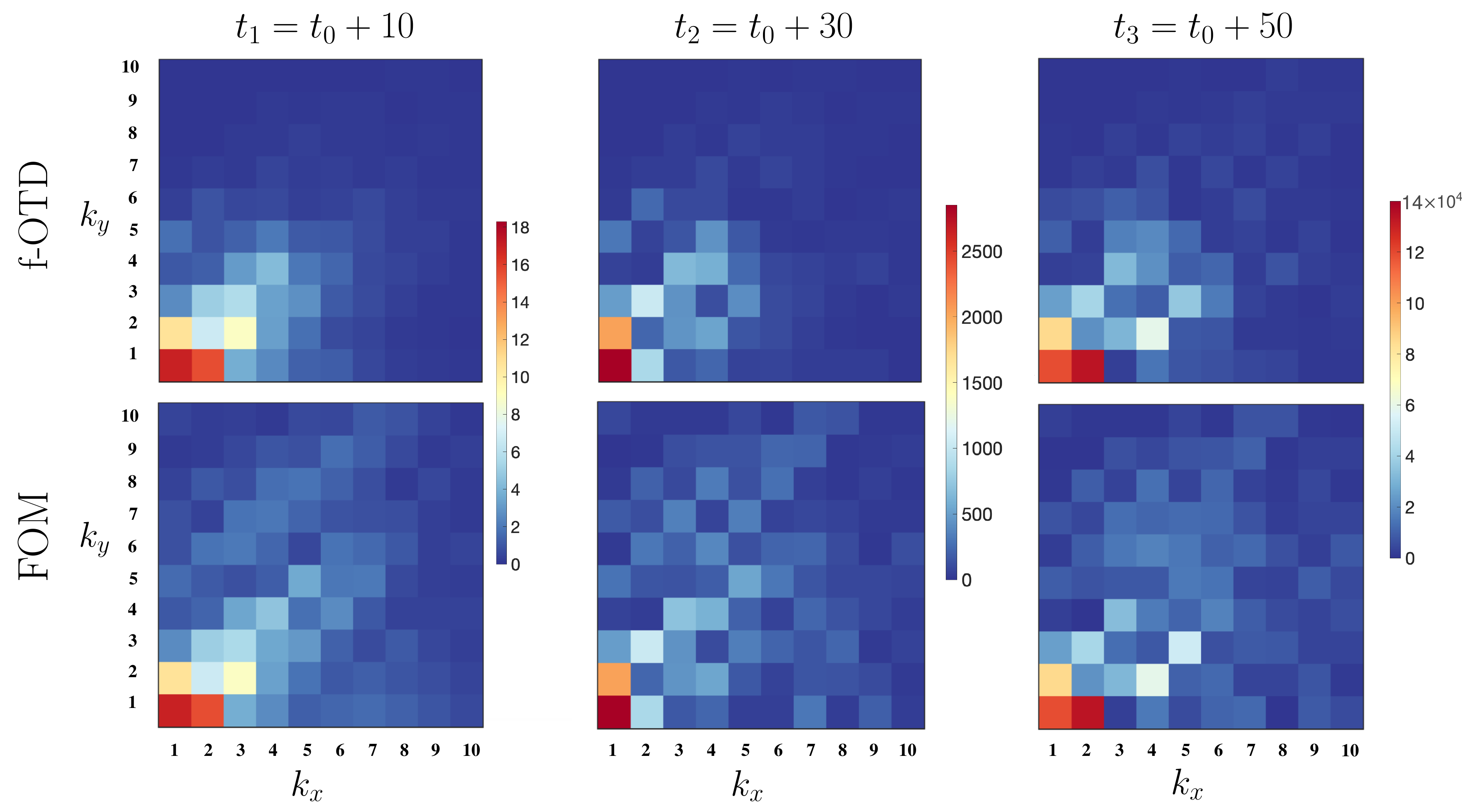}
\caption{The Kolmogorov flow: response ratio for $d=100$ and $r=20$.}
\label{fig:ampli_ratio}
\end{figure}

The f-OTD reconstruction can be used to compute the response ratio for any of the forces. For this problem, the forcing is not time-dependent, and the response ratio is the ratio of the norm of the disturbance to the norm of the corresponding forcing:
\begin{equation}\label{eq:amp_force}
    R_i(t) = \frac{\| \mathbf v_i(x,y,t)\|}{\|\mathbf f_i(x,y)\|}, \quad i=1,2, \dots, d.
\end{equation}
An advantage of computing the response ratio is that it removes the effect of the forcing spectrum; in other words, the forcings that are attenuated in the spectrum are restored to their original magnitudes. If the forcing spectrum is not used, the singular values will cluster near the initial evolution, requiring a large rank.      Therefore, $R_i(t)$ is an objective metric to assess the performance of f-OTD and also an important quantity for analyzing how the flow responds to individual forces.

The norm of $\mathbf v_i$ using f-OTD does not require explicitly forming $\mathbf v_i$ and it can be computed using the low-rank approximation as shown below:
\begin{align*}
\|\mathbf v_i \|^2 = \| \sum_{m=1}^r y_{im} \mathbf u_m  \|^2 &= \inner{\sum_{m=1}^r  y_{im} \mathbf u_m}{\sum_{m'=1}^r y_{im'} \mathbf u_{m'} } 
                                                             = \sum_{m=1}^r \sum_{m'=1}^r y_{im}  y_{im'}\inner{ \mathbf u_m }{ \mathbf u_{m'}}\\
                                                             &= \sum_{m=1}^r \sum_{m'=1}^r y_{im}  y_{im'}\delta_{m m'}
                                                             = \sum_{m=1}^r y^2_{im},
\end{align*} 
where $y_{im}$ is the $(i,m)$ entry of the f-OTD coefficient matrix, i.e., $y_{im} = \mathbf Y_{(i,m)}$. In the above derivation, we have made use of the orthonormality properties of the f-OTD modes. Therefore,  $\| \mathbf v_i \|$ is equal to the second norm of $i$-th row of the $\mathbf Y$ matrix, which can be computed at a negligible computational cost.

The response ratios for the case of $d=100$ and the f-OTD rank of $r=20$ at three different time instants are shown in Figure \ref{fig:ampli_ratio}. These response ratios for 100 forcings are shown in the $k_x$-$k_y$ plane. The corresponding quantities are also calculated from the FOM.  It is clear the flow is much more responsive to the forces with lower wavenumbers. Moreover, comparing the f-OTD results with the FOM shows that the f-OTD approximation tends to be more accurate for the most excited forces. This is not surprising, as f-OTD seeks to capture the most amplified subspace, which in this problem is dominated by response to forces with lower wavenumbers.

Now, we consider the case with $d=16,384$. Solving the FOM for this case is cost-prohibitive. Therefore, we only perform the f-OTD low-rank approximation for three ranks: $r=10, 15$, and 20, to demonstrate convergence.  We show how the f-OTD low-rank approximation can be used to extract quantities that describe the intermittent events. To this end, we compute the eigenvalues of the symmetric part of $\mathbf L_r$. The matrix $\mathbf L_r$ is the reduced linear operator that is obtained by projecting the full-dimensional instantaneous linear operator onto the f-OTD subspace, i.e., $\mathbf L_r = \mathbf U^T \mathbf L \mathbf U \in \mathbb{R}^{r\times r}$. The symmetric part of $\mathbf L_r$ is given by:
\begin{equation}
    \mathbf S_r = \frac{\mathbf L_r+\mathbf L_r^\mathrm{T}}{2}.
\end{equation}
The eigenvalues of $\mathbf S_r$ are real and can be sorted such that $\lambda^S_1 \geq \lambda^S_2 \geq \dots \geq \lambda^S_r$ and $\lambda^S_1$ represents the maximum instantaneous growth rate within the f-OTD subspace (\cite{FIII96}). In Figure \ref{fig:Lr_diss}b, the energy dissipation is depicted which shows an intermittent event during the time interval of $t \in [210, 260]$ as shown in the top and middle panels.  In Figure \ref{fig:Lr_diss}c,  $\lambda^S_1(t)$ for three f-OTD ranks are shown. We observe that the intermittent event causes a significant increase in $\lambda^S_1$. Moreover, as $r$ increases, $\lambda^S_1$ converges to an upper envelope.

\begin{figure}
\centering
\includegraphics[width=0.63\textwidth]{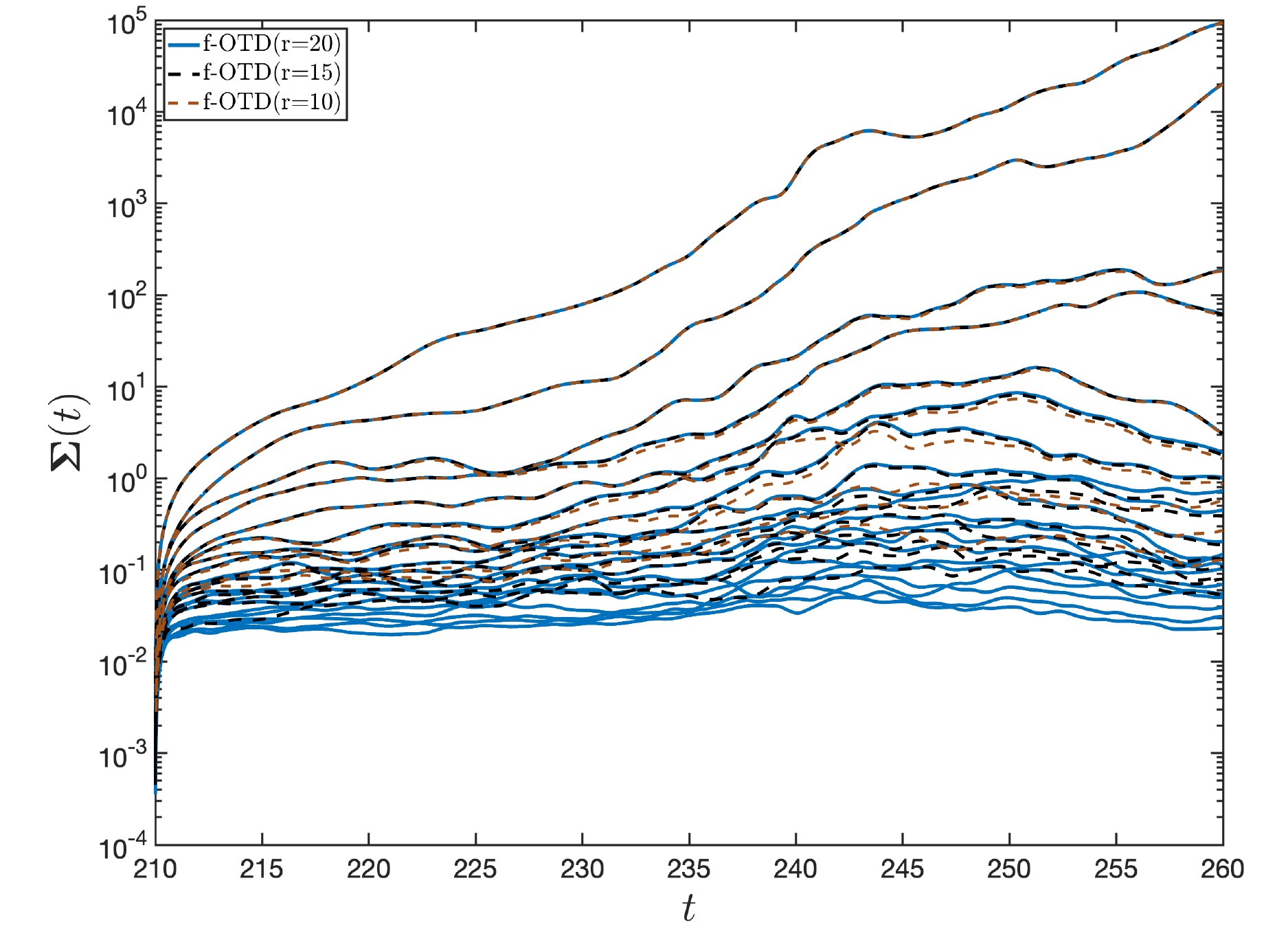}
\caption{The Kolmogorov flow: convergence study for singular values evolution from f-OTD for different reduction sizes ($r=20$, $r=15$, and $r=10$) for $d=16,384$.  }
  \label{fig:sig_kom}
\end{figure}

The singular values for $r=10, 15$, and $20$  are shown in Figure \ref{fig:sig_kom}. It is clear the leading singular values have converged and there is a larger discrepancy between the singular values of $r=10,15$ and 20 for modes with lower energy. The leading singular values grow exponentially with time due to the chaotic nature of the dynamics. There is also a large gap between the leading singular values indicating that the resulting MDE is instantaneously low-rank.

\subsection{Two-dimensional decaying isotropic turbulence} 

In this section, we present the application of f-OTD in low-rank approximation of the response of the linearized Navier-Stokes equation to a large number of impulses, which has utility in flow control and in particular balanced truncation reduced-order models (\cite{M81,LMG99,R05}). 

The post-transient response of the linearized Navier-Stokes equations and their adjoint to an impulse is widely used in control to estimate the Gramians, as solving the time-dependent Lyapunov equation in the time domain is prohibitive for large-scale dynamical systems. When the base flow is steady-state, it is possible to compute the Gramians cost-effectively. For periodic base flows, \cite{PR24} proposed an effective approach to compute the frequential Gramians by solving them in the frequency domain. However, for arbitrarily time-dependent base flows, computing the Gramians requires determining the response of the direct and adjoint linearized Navier-Stokes equations to a very large number of impulses, which is computationally taxing.

In this demonstration, we show that the f-OTD low-rank approximation can reduce the computational cost of solving this problem. Specifically, we illustrate the computational advantages of formulating the response to a large number of impulses as an MDE. We numerically demonstrate that the resulting MDE is instantaneously low-rank. These low-rank properties are then exploited using f-OTD by approximating the solution of the MDE on the manifold of low-rank matrices.

For the demonstration, we consider two-dimensional decaying turbulent flow. This flow is chosen due to its complex dynamics, highly unsteady and time-dependent base flows, and chaotic nature (\cite{jimenez2020monte}). The simulations are conducted using the same numerical methods as those explained in Section \ref{2D_TEJ}. The computational domain is a square with length \( L = 2\pi \) and is discretized with a \( 512 \times 512 \) grid. We use the time step of \(\Delta t =0.001\) for temporal advancement and the $Re=10000$.  The flow fields are initialized using the  Taylor-Green vortex (\cite{vuorinen2016dnslab}).


\begin{figure}
\centering
\includegraphics[width=0.63\textwidth]{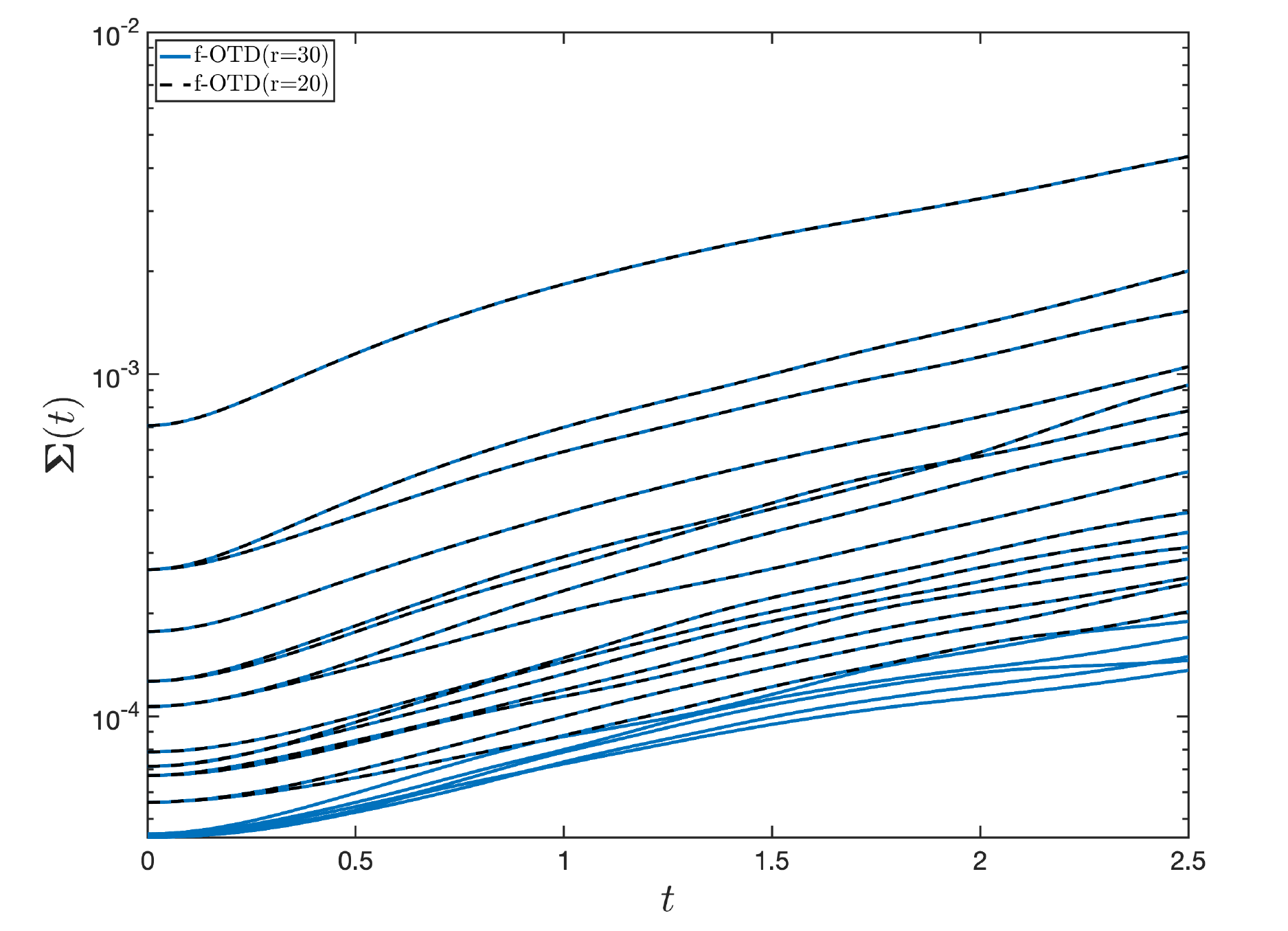}
\caption{2D decaying isotropic turbulence: singular value of f-OTD with $r = 20$ and $r=30$ of post-transient response matrix to $d=1024$ impulses.}
  \label{fig:SVD_2D_Tubulent}
\end{figure}

\begin{figure}
\centering
\includegraphics[width=0.88\textwidth]{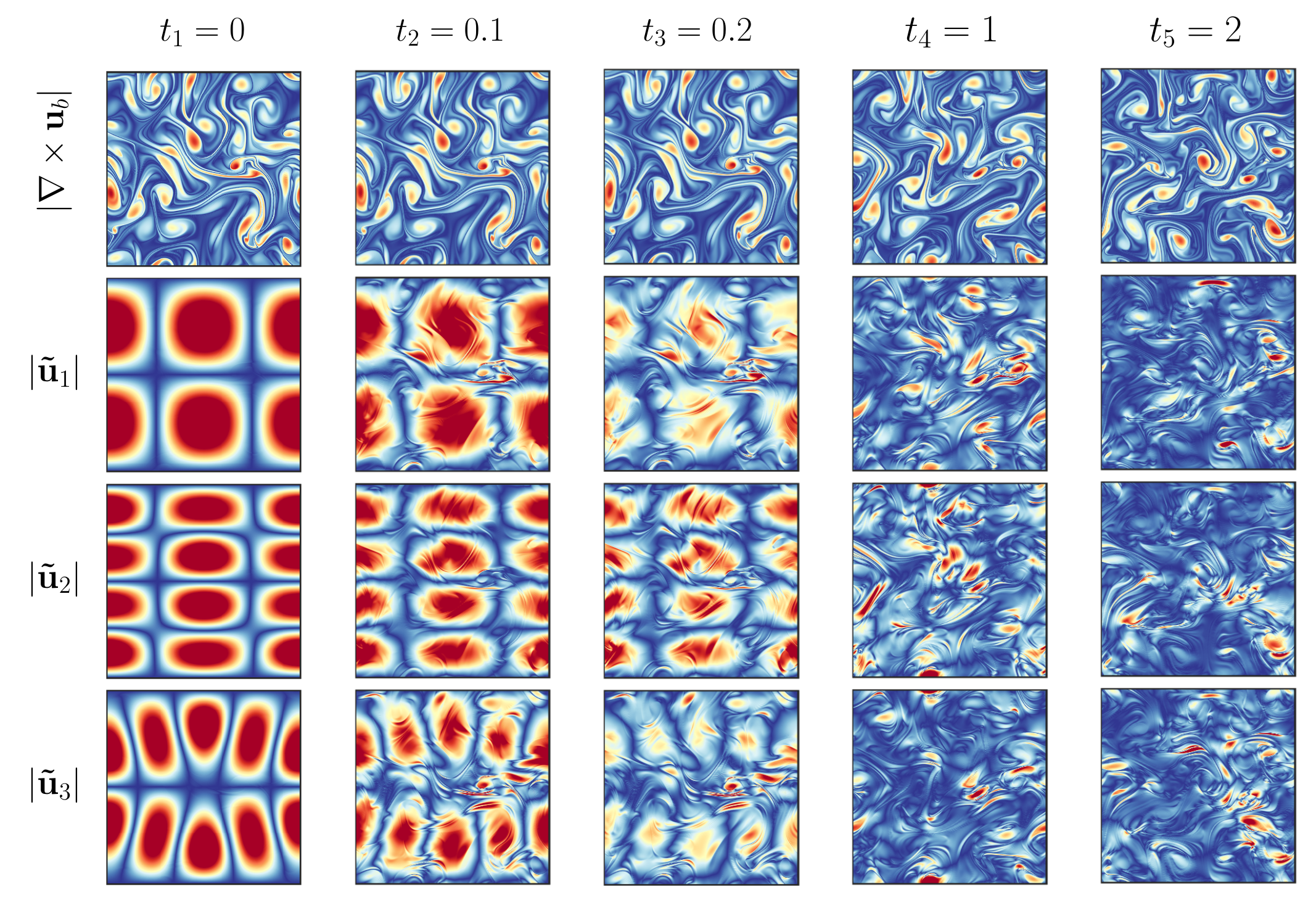}
\caption{2D decaying isotropic turbulence: evolution of the first three dominant f-OTD modes and the base flow for low-rank approximation of the post-transient response to $d=1024$ impulses. The base flow is visualized by the magnitude of vorticity, $|\nabla \times \mathbf u_b|$ and the modes are visualized by the magnitude of f-OTD vector at each grid point, $|\tilde{\mathbf u}_i| = (\tilde{u}_{x_i}^2+\tilde{u}_{y_i}^2)^{1/2}$.  }
  \label{fig:Umode_2D_Tubulent}
\end{figure}

For this problem, we employ an external forcing in space that acts impulsively at the initial time.   We define the external forcing as:

\begin{equation*}
f_{x}(x,y,t)= \frac{c}{k_x }\cos(k_x x) \sin(k_y y) g(t)   \quad \quad 
f_{y}(x,y,t)= - \dfrac{c}{k_y} \sin(k_x x)\cos(k_y y) g(t),
\end{equation*}

where  $g(t)$ is the impulse function and is defined as:
\[ g(t) = \begin{cases} 
1 & \text{if } t = t_0, \\
0 & \text{if } t > t_0,
\end{cases} \]
where $t_0=\Delta t$.

The forcing spectrum is the same as that used for the Kolmogorov flow. In this study, the wave numbers $k_x$ and $k_y$ ,  varied from $1$ to $32$. We considered a total of $d=1024$ external excitations and employed a subspace size of $r=30$ for the f-OTD reduction, which yielded satisfactory results. For this analysis, we only conducted f-OTD simulations. The f-OTD was initialized by computing the FOM for a single time step, after which the initial conditions for the f-OTD were assigned by using the first $r$-rank from the SVD of the FOM solution.

To demonstrate the convergence of the f-OTD modes, we compare the leading singular values from two different subspace sizes, $r=20$ and $r=30$. As illustrated in Figure \ref{fig:SVD_2D_Tubulent}, the singular values match well with each other, indicating that $r=20$ is sufficiently large. There is also a large gap between the singular values, especially between the first singular values and the rest indicating this problem is instantaneously low rank.   

The computational cost of solving the f-OTD equations is roughly equal to solving $r$ forced linearized Navier-Stokes equations. See Section \ref{sec:comp_cost} for more details. Therefore, the computational savings offered by f-OTD,  in terms of floating point operations,  memory, and input/output is roughly equal to the factor of $d/r$. 

Figure  \ref{fig:Umode_2D_Tubulent} presents the temporal evolution of the first three dominant f-OTD modes in canonical form as well as the base flow at five time instances. The base flow is visualized by the values of the voriticty, $|\nabla \times \mathbf u_b|$ and the f-OTD modes, in the canonical form, are visualized by the f-OTD vector at each grid point, i.e., $|\tilde{\mathbf u}_i(x,y,t)| = (\tilde{u}_{x_i}(x,y,t)^2+\tilde{u}_{y_i}(x,y,t)^2)^{1/2}$, where $\tilde{u}_{x_i}$ and $\tilde{u}_{y_i}$ are the $x$ and $y$ components of $\tilde{\mathbf u}_i(x,y,t)$. The results show a rapid realignment of the initial f-OTD subspace to the dynamically dominant perturbation subspace in the early stages of the evolution ($0<t<0.2$). During this period, although the base flow has barely changed, the f-OTD subspace evolves very quickly. As time progresses, the effect of the base flow evolution on the f-OTD modes becomes evident, resulting in rich and complex modes that evolve with the base flow. The response to the impulse becomes highly localized over time, as manifested by the strong presence of the f-OTD modes in small regions of the flow.

%% file: Conclusions.tex
\section{ Conclusion \label{Conclusion}}
Analysis of linear disturbance growth due to external forcing is crucial for flow stability, control, and uncertainty quantification. However, when the base flow is arbitrarily time-dependent, the computational tools effective for analyzing steady-state base flows become either inadequate or too computationally prohibitive to use. To this end, we present a methodology to build low-rank solution operators for the linear evolution of disturbance for problems with unsteady base flows. The solution operator is a time-dependent matrix whose evolution equation is governed by forced linearized dynamics.    The formulation is based on the forced optimally time-dependent decomposition (f-OTD), in which the solution operator is approximated by the multiplication of two skinny time-dependent matrices.  Using a variational principle, evolution equations for these two matrices are derived. The f-OTD low-rank approximation is equivalent to the dynamical low-rank approximation and similarly constrains the solution of the matrix differential equation to a manifold of low-rank matrices. 

We demonstrate the utility of the developed methodology through several case studies. These demonstrations show how the methodology can identify the optimal forcing and employ the operator as an effective surrogate model.   We also show the connection between the presented methodology and the resolvent analysis. In particular, we numerically  show when applied to the steady-state mean flow,  the presented low-rank approximation asymptotically converges to that of the resolvent analysis. 

Two mathematical developments are of interest for future studies: (i) rigorously proving the connection between the asymptotic behavior of f-OTD and resolvent analysis; (ii) developing a rank-adaptive f-OTD methodology that determines the rank of the low-rank approximation at each time instant, based on the accuracy requirements of the solution operator. 
\section*{Acknowledgements}
We gratefully acknowledge the support funding from Transformational Tools and Technology (TTT), NASA grant no. 80NSSC22M0282. and the National Science Foundation (NSF), USA, under the Grant CBET2042918.
This research was supported in part by the University of Pittsburgh
Center for Research Computing through the resources provided.

%% file: Appendix.tex
\appendix
\section{f-OTD derivation for incompressible Navier–Stokes}
\label{Appendix_f-OTD_NS}
In this section, we present the f-OTD evolution equations for the incompressible Navier-Stokes equations. The  incompressible Navier-Stokes equations govern the evolution of the base flow as shown below:
\begin{align}
\frac{\partial {U_b}}{\partial t} + ({U_b}\cdot ){\nabla U_b} &= -{\nabla p}_b + \frac{1}{Re}\nabla^{2}{U_b}, \label{eq:NS_u}\\
{\nabla}\cdot {U_b} &= 0, \label{eq:NS_v}
\end{align}
where ${U_b}(x,y,t)$ is the velocity vector field with components ${U_b}(x,y,t) = (u_b(x,y,t), v_b(x,y,t))$, and ${p}_b(x,y,t)$ is the pressure field. The  evolution  equation for disturbance $v_i(x,y,t)$ is given by
\begin{align}
\frac{\partial v_i}{\partial t}  &= \mathcal{L}_{NS}(v_i)\label{eq:LNS_u}-\nabla p_i + f_i\\
{\nabla}\cdot {v_i} &= 0, \label{eq:LNS_div}
\end{align}
where $\mathcal{L}_{NS}(\sim)$ is the linearized Navier-Stokes operator  given by
\begin{equation}\label{eq:L_NS}
    \mathcal{L}_{NS}(v_i) = - (U_b \cdot \nabla)v_i - ( v_i \cdot \nabla)U_b+\frac{1}{Re} \nabla^2 v_i, \quad i=1,2,\dots, d
\end{equation}
and $f_i(x,y,t)$ is the external forcing.  The matrix $\mathbf{L}$ and vector $\mathbf{f}_i$ in the f-OTD evolution equations are the discrete representation of $\mathcal{L}_{NS}$ and $f_i(x,y,t)$, respectively.  
 For incompressible Navier-Stokes equation, each f-OTD mode is a vector field, that is, \( u_i = (u_{x_i}, u_{y_i}) \), for \( i=1, \ldots, r \). The f-OTD modes are orthonormal with respect to the following inner product:
\begin{equation*}
\inner{u_i}{u_j}  = \int \int (u_{x_i} u_{x_j} + u_{y_i} u_{y_j}) \, dx \, dy, \quad \quad \|u \| = \inner{u}{u}^{1/2}.
\end{equation*}
The induced norm is also defined above.
Moreover,  the f-OTD vector field must be divergence-free to ensure that any f-OTD reconstructed field is also divergence-free. Therefore, $\nabla \cdot u_i =0$.  To enforce the divergence-free condition, we use the projection method. Below, we explain the steps for the time integration for the f-OTD equations for an explicit Euler scheme:
\begin{enumerate}
\item Solve Eq. \ref{eq:u_ev_DO}, in  which $\mathbf{L}$ is the discrete representation of $\mathcal{L}_{NS}$ operator given by Eq. \ref{eq:L_NS}. 
\begin{equation}
\hat{\mathbf{U}}^{k+1} = \mathbf{U}^{k} + \Delta t \big (\mathbf{L}^k\mathbf{U}^k-\mathbf{U}^k\mathbf{L}^k_r+\big(\mathbf{F}^k\mathbf{Y}^k-\mathbf{U}^k\mathbf{U}^{k^T}\mathbf{F}^k\mathbf{Y}^k\big){\mathbf{C}}^{k^{-1}} \big),
\end{equation}
where the solution at time step $k$ is denoted with a superscript $(\sim) ^k$ and $\Delta t$ is the time advancement magnitude. For an efficient implementation of the above step, it is important to note that the matrix $\mathbf{L}$ does not need to be explicitly formed; rather, what is required is the action of $\mathbf{L}$ on vector fields. 
\item  Note that the pressure gradient is not included in the linearized operator, and as a result, the columns of   $\hat{\mathbf{U}}^{k+1}$, denoted by $\hat{\mathbf{u}}^{k+1}_i$, are not divergence-free. In this step, we use the projection method to project $\hat{\mathbf{u}}^{k+1}_i$ onto a divergence-free vector field:  

\begin{align*}
   \mathbf{D} \mathbf{G} \boldsymbol{\phi}^{k+1}_i &= \mathbf{D} \hat{\mathbf{u}}^{k+1}_i, \quad i=1,2, \dots, r,\\
   \mathbf{u}_i^{k+1} &= \hat{\mathbf{u}}^{k+1}_i - \mathbf{G} \boldsymbol{\phi}^{k+1}_i
\end{align*}
where $\mathbf{D} \in \mathbb{R}^{n \times 2n }$ and  $\mathbf{G} \in \mathbb{R}^{2n \times n}$ are the discrete representations of the gradient and divergence operators, respectively. 
\item In the last step, we advance the evolution equation for the f-OTD coefficients using the divergence-free f-OTD modes: 
\begin{equation}
\mathbf{Y}^{k+1} = \mathbf{Y}^k + \Delta t \big(  \mathbf{Y}^k\mathbf{L}_r^{k^T}+\mathbf{F}^{k^T}\mathbf{U}^k \big).
\end{equation}
\end{enumerate}
 We use the fourth-order Runge-Kutta scheme for the time integration of the f-OTD equations. Time advancement of every stage of the Runge-Kutta scheme is analogous to the explicit Euler explained above.

%% file: main.bbl
\begin{thebibliography}{10}

\bibitem{schmid2007nonmodal}
P.~J. Schmid.
\newblock Nonmodal stability theory.
\newblock {\em Annu. Rev. Fluid Mech.}, 39:129--162, 2007.

\bibitem{qin2016response}
Fufeng Qin and Xuesong Wu.
\newblock Response and receptivity of the hypersonic boundary layer past a wedge to free-stream acoustic, vortical and entropy disturbances.
\newblock {\em Journal of Fluid Mechanics}, 797:874--915, 2016.

\bibitem{schrader2009receptivity}
Lars-Uve Schrader, Luca Brandt, and Dan~S Henningson.
\newblock Receptivity mechanisms in three-dimensional boundary-layer flows.
\newblock {\em Journal of Fluid Mechanics}, 618:209--241, 2009.

\bibitem{schneider2001effects}
Steven~P Schneider.
\newblock Effects of high-speed tunnel noise on laminar-turbulent transition.
\newblock {\em Journal of Spacecraft and Rockets}, 38(3):323--333, 2001.

\bibitem{joo2012continuous}
Jongwook Joo and Paul~A Durbin.
\newblock Continuous mode transition in high-speed boundary-layers.
\newblock {\em Flow, turbulence and combustion}, 88(3):407--430, 2012.

\bibitem{brinkerhoff2015numerical}
Joshua~R Brinkerhoff and Metin~I Yaras.
\newblock Numerical investigation of transition in a boundary layer subjected to favourable and adverse streamwise pressure gradients and elevated free stream turbulence.
\newblock {\em Journal of Fluid Mechanics}, 781:52--86, 2015.

\bibitem{delsole1995stochastically}
Timothy DelSole and Brian~F Farrell.
\newblock A stochastically excited linear system as a model for quasigeostrophic turbulence: Analytic results for one-and two-layer fluids.
\newblock {\em Journal of Atmospheric Sciences}, 52(14):2531--2547, 1995.

\bibitem{KLC23}
O.~Kamal, M.~T. Lakebrink, and T.~Colonius.
\newblock Global receptivity analysis: physically realizable input--output analysis.
\newblock {\em Journal of Fluid Mechanics}, 956:R5, 2023.

\bibitem{edwards1994krylov}
WS~Edwards, Laurette~S Tuckerman, Richard~A Friesner, and DC1259901 Sorensen.
\newblock Krylov methods for the incompressible navier-stokes equations.
\newblock {\em Journal of computational physics}, 110(1):82--102, 1994.

\bibitem{frantz2023krylov}
Ricardo~AS Frantz, J-Ch Loiseau, and J-Ch Robinet.
\newblock Krylov methods for large-scale dynamical systems: Application in fluid dynamics.
\newblock {\em Applied Mechanics Reviews}, 75(3):030802, 2023.

\bibitem{bagheri2009global}
S.~Bagheri, P.~Schlatter, P.~J. Schmid, and D.~S. Henningson.
\newblock Global stability of a jet in crossflow.
\newblock {\em Journal of Fluid Mechanics}, 624:33--44, 2009.

\bibitem{aakervik2006steady}
E.~{\AA}kervik, L.~Brandt, D.S. Henningson, J.~H{\oe}pffner, O.~Marxen, and P.~Schlatter.
\newblock Steady solutions of the navier-stokes equations by selective frequency damping.
\newblock {\em Physics of Fluids}, 18:068102, 2006.

\bibitem{Henningson_Schmidt}
P.~J. Schmid and D.~S. Henningson.
\newblock {\em Stability and Transition Stability in Shear Flows}.
\newblock Springer, 2001.

\bibitem{trefethen1993hydrodynamic}
Lloyd~N Trefethen, Anne~E Trefethen, Satish~C Reddy, and Tobin~A Driscoll.
\newblock Hydrodynamic stability without eigenvalues.
\newblock {\em Science}, 261(5121):578--584, 1993.

\bibitem{schmidt2018spectral}
Oliver~T Schmidt, Aaron Towne, Georgios Rigas, Tim Colonius, and Guillaume~A Bres.
\newblock Spectral analysis of jet turbulence.
\newblock {\em Journal of Fluid Mechanics}, 855:953--982, 2018.

\bibitem{bae2020resolvent}
H~Jane Bae, Scott~TM Dawson, and Beverley~J McKeon.
\newblock Resolvent-based study of compressibility effects on supersonic turbulent boundary layers.
\newblock {\em Journal of Fluid Mechanics}, 883:A29, 2020.

\bibitem{skene2022sparsifying}
Calum~S Skene, Chi-An Yeh, Peter~J Schmid, and Kunihiko Taira.
\newblock Sparsifying the resolvent forcing mode via gradient-based optimisation.
\newblock {\em Journal of Fluid Mechanics}, 944:A52, 2022.

\bibitem{cook2022free}
David~A Cook and Joseph~W Nichols.
\newblock Free-stream receptivity of a hypersonic blunt cone using input--output analysis and a shock-kinematic boundary condition.
\newblock {\em Theoretical and Computational Fluid Dynamics}, 36(1):155--180, 2022.

\bibitem{mckeon2010critical}
Beverley~J McKeon and Ati~S Sharma.
\newblock A critical-layer framework for turbulent pipe flow.
\newblock {\em Journal of Fluid Mechanics}, 658:336--382, 2010.

\bibitem{abreu2020resolvent}
Leandra~I Abreu, Andr{\'e}~VG Cavalieri, Philipp Schlatter, Ricardo Vinuesa, and Dan~S Henningson.
\newblock Resolvent modeling of near-wall coherent structures in turbulent channel flow.
\newblock {\em International Journal of Heat and Fluid Flow}, 85:108662, 2020.

\bibitem{martini2022resolvent}
Eduardo Martini, Junoh Jung, Andr{\'e}~VG Cavalieri, Peter Jordan, and Aaron Towne.
\newblock Resolvent-based tools for optimal estimation and control via the {Wiener}--{Hopf} formalism.
\newblock {\em Journal of Fluid Mechanics}, 937:A19, 2022.

\bibitem{jin2022optimal}
Bo~Jin, Simon~J Illingworth, and Richard~D Sandberg.
\newblock Optimal sensor and actuator placement for feedback control of vortex shedding.
\newblock {\em Journal of Fluid Mechanics}, 932:A2, 2022.

\bibitem{mckeon2017engine}
BJ~McKeon.
\newblock The engine behind (wall) turbulence: perspectives on scale interactions.
\newblock {\em Journal of Fluid Mechanics}, 817:P1, 2017.

\bibitem{sun2020resolvent}
Yiyang Sun, Qiong Liu, Louis~N Cattafesta~III, Lawrence~S Ukeiley, and Kunihiko Taira.
\newblock Resolvent analysis of compressible laminar and turbulent cavity flows.
\newblock {\em AIAA journal}, 58(3):1046--1055, 2020.

\bibitem{ribeiro2020randomized}
Jean H{\'e}lder~Marques Ribeiro, Chi-An Yeh, and Kunihiko Taira.
\newblock Randomized resolvent analysis.
\newblock {\em Physical Review Fluids}, 5(3):033902, 2020.

\bibitem{MAB10}
A.~Monokrousos, E.~{\AA}kervik, L.~Brandt, and D.~S. Henningson.
\newblock Global three-dimensional optimal disturbances in the blasius boundary-layer flow using time-steppers.
\newblock 650:181--214, 2010.

\bibitem{MRTC21}
E.~Martini, D.~Rodr{\'\i}guez, A.~Towne, and A.~G. Cavalieri.
\newblock Efficient computation of global resolvent modes.
\newblock 919:A3, 2021.

\bibitem{L2003}
James~Hetao Liu.
\newblock {\em A First Course in the Qualitative Theory of Differential Equations}.
\newblock Prentice Hall, Upper Saddle River, 2003.

\bibitem{KLH23}
J.~S. Kern, V.~Lupi, and D.~S. Henningson.
\newblock Floquet stability analysis of pulsatile flow in toroidal pipes.
\newblock {\em Physical Review Fluids}, 9(4):043906--, 04 2024.

\bibitem{padovan2020analysis}
Alberto Padovan, Samuel~E Otto, and Clarence~W Rowley.
\newblock Analysis of amplification mechanisms and cross-frequency interactions in nonlinear flows via the harmonic resolvent.
\newblock {\em Journal of Fluid Mechanics}, 900, 2020.

\bibitem{FJBT24}
A.~Farghadan, J.~Jung, R.~Bhagwat, and A.~Towne.
\newblock Efficient harmonic resolvent analysis via time stepping.
\newblock {\em Theoretical and Computational Fluid Dynamics}, 2024.

\bibitem{lopez2023sparsity}
Barbara Lopez-Doriga, Eric Ballouz, Hyunji~Jane Bae, and Scott~T Dawson.
\newblock A sparsity-promoting resolvent analysis for the identification of spatiotemporally-localized amplification mechanisms.
\newblock In {\em AIAA SCITECH 2023 Forum}, page 0677, 2023.

\bibitem{KNH24}
J.~S. Kern, P.~S. Negi, A.~Hanifi, and D.~S. Henningson.
\newblock Onset of absolute instability on a pitching aerofoil.
\newblock 988:A8, 2024.

\bibitem{T21}
T.~P. Sapsis.
\newblock Statistics of extreme events in fluid flows and waves.
\newblock {\em Annual Review of Fluid Mechanics}, 53(Volume 53, 2021):85--111, 2021.

\bibitem{AML11}
K.~Avila, D.~Moxey, A.~D. Lozar, M.~Avila, D.~Barkley, and B.~Hof.
\newblock The onset of turbulence in pipe flow.
\newblock {\em Science}, 333(6039):192--196, 2011.

\bibitem{CD13}
S.~Cherubini and P.~De~Palma.
\newblock Nonlinear optimal perturbations in a couette flow: bursting and transition.
\newblock 716:251--279, 2013.

\bibitem{FCRP17}
Mirko Farano, Stefania Cherubini, Jean-Christophe Robinet, and Pietro De~Palma.
\newblock Optimal bursts in turbulent channel flow.
\newblock 817:35--60, 2017.

\bibitem{SS19}
O.~T. Schmidt and P.~J. Schmid.
\newblock A conditional space--time pod formalism for intermittent and rare events: example of acoustic bursts in turbulent jets.
\newblock 867:R2, 2019.

\bibitem{MCT11}
Y.~Modarres-Sadeghi, F.~Chasparis, M.~S. Triantafyllou, M.~Tognarelli, and P.~Beynet.
\newblock Chaotic response is a generic feature of vortex-induced vibrations of flexible risers.
\newblock {\em Journal of Sound and Vibration}, 330(11):2565--2579, 2011.

\bibitem{DCG12}
P.~Davini, C.~Cagnazzo, S.~Gualdi, and A.~Navarra.
\newblock Bidimensional diagnostics, variability, and trends of northern hemisphere blocking.
\newblock {\em Journal of Climate}, 25(19):6496--6509, 2012.

\bibitem{donello2022computing}
Michael Donello, Mark~H Carpenter, and Hessam Babaee.
\newblock Computing sensitivities in evolutionary systems: a real-time reduced order modeling strategy.
\newblock {\em SIAM Journal on Scientific Computing}, 44(1):A128--A149, 2022.

\bibitem{babaee2016minimization}
H~Babaee and TP~Sapsis.
\newblock A minimization principle for the description of modes associated with finite-time instabilities.
\newblock {\em Proceedings of the Royal Society A: Mathematical, Physical and Engineering Sciences}, 472(2186):20150779, 2016.

\bibitem{babaee2017reduced}
Hessam Babaee, Mohamad Farazmand, George Haller, and Themistoklis~P Sapsis.
\newblock Reduced-order description of transient instabilities and computation of finite-time lyapunov exponents.
\newblock {\em Chaos: An Interdisciplinary Journal of Nonlinear Science}, 27(6):063103, 2017.

\bibitem{beneitez2020edge}
Miguel Beneitez, Yohann Duguet, Philipp Schlatter, and Dan~S Henningson.
\newblock Edge manifold as a {L}agrangian coherent structure in a high-dimensional state space.
\newblock {\em Physical Review Research}, 2(3):033258, 2020.

\bibitem{FS16}
M.~Farazmand and T.~P. Sapsis.
\newblock Dynamical indicators for the prediction of bursting phenomena in high-dimensional systems.
\newblock {\em Phys. Rev. E}, 94:032212, Sep 2016.

\bibitem{blanchard2019stabilization}
Antoine Blanchard and Themistoklis~P Sapsis.
\newblock Stabilization of unsteady flows by reduced-order control with optimally time-dependent modes.
\newblock {\em Physical Review Fluids}, 4(5):053902, 2019.

\bibitem{kern2021transient}
J~Simon Kern, Miguel Beneitez, Ardeshir Hanifi, and Dan~S Henningson.
\newblock Transient linear stability of pulsating {P}oiseuille flow using optimally time-dependent modes.
\newblock {\em Journal of Fluid Mechanics}, 927, 2021.

\bibitem{nouri2022skeletal}
AG~Nouri, H~Babaee, P~Givi, HK~Chelliah, and Daniel Livescu.
\newblock Skeletal model reduction with forced optimally time dependent modes.
\newblock {\em Combustion and Flame}, 235:111684, 2022.

\bibitem{nouri2024skeletal}
AG~Nouri, Y~Liu, P~Givi, H~Babaee, and D~Livescu.
\newblock Skeletal kinetics reduction for astrophysical reaction networks.
\newblock {\em The Astrophysical Journal Supplement Series}, 272(2):34, 2024.

\bibitem{Beck:2000aa}
M.~H. Beck, A.~J{\"a}ckle, G.~A. Worth, and H.~D. Meyer.
\newblock The multiconfiguration time-dependent {H}artree ({MCTDH}) method: a highly efficient algorithm for propagating wavepackets.
\newblock {\em Physics Reports}, 324(1):1--105, 1 2000.

\bibitem{KL07}
O.~Koch and C.~Lubich.
\newblock Dynamical low‐rank approximation.
\newblock {\em SIAM Journal on Matrix Analysis and Applications}, 29(2):434--454, 2017/04/02 2007.

\bibitem{SL09}
T.P. Sapsis and P.F.J. Lermusiaux.
\newblock Dynamically orthogonal field equations for continuous stochastic dynamical systems.
\newblock {\em Physica D: Nonlinear Phenomena}, 238(23-24):2347--2360, 2009.

\bibitem{PB20}
P.~Patil and H.~Babaee.
\newblock Real-time reduced-order modeling of stochastic partial differential equations via time-dependent subspaces.
\newblock {\em Journal of Computational Physics}, 415:109511, 2020.

\bibitem{RNB21}
D.~Ramezanian, A.~G. Nouri, and H.~Babaee.
\newblock On-the-fly reduced order modeling of passive and reactive species via time-dependent manifolds.
\newblock {\em Computer Methods in Applied Mechanics and Engineering}, 382:113882, 2021.

\bibitem{einkemmer2018low}
Lukas Einkemmer and Christian Lubich.
\newblock A low-rank projector-splitting integrator for the {V}lasov--{P}oisson equation.
\newblock {\em SIAM Journal on Scientific Computing}, 40(5):B1330--B1360, 2018.

\bibitem{KS23}
{Kusch, J.} and {Stammer, P.}
\newblock A robust collision source method for rank adaptive dynamical low-rank approximation in radiation therapy.
\newblock {\em ESAIM: M2AN}, 57(2):865--891, 2023.

\bibitem{TSC18}
A.~Towne, O.~T. Schmidt, and .~Colonius.
\newblock Spectral proper orthogonal decomposition and its relationship to dynamic mode decomposition and resolvent analysis.
\newblock 847:821--867, 2018.

\bibitem{FT23}
P.~Frame and A.~Towne.
\newblock Space-time {POD} and the {H}ankel matrix.
\newblock {\em PLOS ONE}, 18(8):e0289637--, 08 2023.

\bibitem{PR24}
A.~Padovan and C.~W. Rowley.
\newblock Continuous-time balanced truncation for time-periodic fluid flows using frequential gramians.
\newblock {\em Journal of Computational Physics}, 496:112597, 2024.

\bibitem{R05}
C.~W. Rowley.
\newblock Model reduction for fluids, using balanced proper orthogonal decomposition.
\newblock {\em International Journal of Bifurcation and Chaos}, 15(03):997--1013, 2005.

\bibitem{DPNFB23}
M.~Donello, G.~Palkar, M.~H. Naderi, D.~C. Del Rey~Fern{\'a}ndez, and H.~Babaee.
\newblock Oblique projection for scalable rank-adaptive reduced-order modelling of nonlinear stochastic partial differential equations with time-dependent bases.
\newblock {\em Proceedings of the Royal Society A: Mathematical, Physical and Engineering Sciences}, 479(2278):20230320, 2023/10/19 2023.

\bibitem{GB24-TT}
B.~Ghahremani and H.~Babaee.
\newblock Cross interpolation for solving high-dimensional dynamical systems on low-rank tucker and tensor train manifolds, 2024.

\bibitem{DRV21}
A.~Dektor, A.~Rodgers, and D.~Venturi.
\newblock Rank-adaptive tensor methods for high-dimensional nonlinear pdes.
\newblock {\em Journal of Scientific Computing}, 88(2):36, 2021.

\bibitem{CKL22}
G.~Ceruti, J.~Kusch, and C.~Lubich.
\newblock A rank-adaptive robust integrator for dynamical low-rank approximation.
\newblock {\em BIT Numerical Mathematics}, 62(4):1149--1174, 2022.

\bibitem{LO14}
C.~Lubich and I.~V. Oseledets.
\newblock A projector-splitting integrator for dynamical low-rank approximation.
\newblock {\em BIT Numerical Mathematics}, 54(1):171--188, 2014.

\bibitem{CL21}
G.~Ceruti and C.~Lubich.
\newblock An unconventional robust integrator for dynamical low-rank approximation.
\newblock {\em BIT Numerical Mathematics}, 2021.

\bibitem{NB23}
M.~H. Naderi and H.~Babaee.
\newblock Adaptive sparse interpolation for accelerating nonlinear stochastic reduced-order modeling with time-dependent bases.
\newblock {\em Computer Methods in Applied Mechanics and Engineering}, 405:115813, 2023.

\bibitem{CHZI13}
M.~Cheng, T.~Y. Hou, and Z.~Zhang.
\newblock A dynamically bi-orthogonal method for time-dependent stochastic partial differential equations i: Derivation and algorithms.
\newblock {\em Journal of Computational Physics}, 242(0):843 -- 868, 2013.

\bibitem{CSK14}
M.~Choi, T.~P. Sapsis, and G.~E. Karniadakis.
\newblock On the equivalence of dynamically orthogonal and bi-orthogonal methods: Theory and numerical simulations.
\newblock {\em Journal of Computational Physics}, 270:1 -- 20, 2014.

\bibitem{BCSK17}
H.~Babaee, M.~Choi, T.~P. Sapsis, and G.~E. Karniadakis.
\newblock A robust bi-orthogonal/dynamically-orthogonal method using the covariance pseudo-inverse with application to stochastic flow problems.
\newblock {\em Journal of Computational Physics}, 344:303--319, 9 2017.

\bibitem{noack2003hierarchy}
B.R. Noack, K.~Afanasiev, M.~Morzynski, G.~Tadmor, and F.~Thiele.
\newblock A hierarchy of low-dimensional models for the transient and post-transient cylinder wake.
\newblock {\em Journal of Fluid Mechanics}, 497(-1):335--363, 2003.

\bibitem{kassam2005fourth}
Aly-Khan Kassam and Lloyd~N Trefethen.
\newblock Fourth-order time-stepping for stiff {PDE}s.
\newblock {\em SIAM Journal on Scientific Computing}, 26(4):1214--1233, 2005.

\bibitem{broze1994nonlinear}
George Broze and Fazle Hussain.
\newblock Nonlinear dynamics of forced transitional jets: temporal attractors and transitions to chaos.
\newblock In {\em Nonlinear Instability of Nonparallel Flows}, pages 459--473. Springer, 1994.

\bibitem{arratia2013transient}
Cristobal Arratia, CP~Caulfield, and J-M Chomaz.
\newblock Transient perturbation growth in time-dependent mixing layers.
\newblock {\em Journal of Fluid Mechanics}, 717:90--133, 2013.

\bibitem{PSF91}
N.~Platt, L.~Sirovich, and N.~Fitzmaurice.
\newblock {An investigation of chaotic Kolmogorov flows}.
\newblock {\em Physics of Fluids A: Fluid Dynamics}, 3(4):681--696, 04 1991.

\bibitem{F18}
E.~D. Fylladitakis.
\newblock Kolmogorov flow: Seven decades of history.
\newblock {\em Journal of Applied Mathematics and Physics}, 6(11):2227--2263, 2018.

\bibitem{FIII96}
B.~F. Farrell and P.~J. Ioannou.
\newblock Generalized stability theory. {P}art {II}: Nonautonomous operators.
\newblock {\em Journal of the Atmospheric Sciences}, 53(14):2041--2053, 2015/04/01 1996.

\bibitem{M81}
B.~Moore.
\newblock Principal component analysis in linear systems: Controllability, observability, and model reduction.
\newblock {\em IEEE Transactions on Automatic Control}, 26(1):17--32, 1981.

\bibitem{LMG99}
S.~Lall, J.E. Marsden, and S.~Glava{\v s}ki.
\newblock Empirical model reduction of controlled nonlinear systems.
\newblock {\em IFAC Proceedings Volumes}, 32(2):2598--2603, 1999.

\bibitem{jimenez2020monte}
Javier Jim{\'e}nez.
\newblock Monte {C}arlo science.
\newblock {\em Journal of Turbulence}, 21(9-10):544--566, 2020.

\bibitem{vuorinen2016dnslab}
Ville Vuorinen and K~Keskinen.
\newblock {DNSL}ab: {A} gateway to turbulent flow simulation in {M}atlab.
\newblock {\em Computer Physics Communications}, 203:278--289, 2016.

\end{thebibliography}
